\documentclass{jfm}

\usepackage{graphicx}
\usepackage{epstopdf, epsfig}
\usepackage{amsmath}
\usepackage{amsfonts}
\usepackage{amssymb}
\usepackage{graphicx}
\usepackage{color}
\usepackage{bm}
\usepackage{xcolor}
\usepackage{caption}
\usepackage{bbold}
\usepackage[T1]{fontenc}
\usepackage{lmodern}

\usepackage{rotating}

\newcommand*{\colorboxed}{}
\def\colorboxed#1#{%
  \colorboxedAux{#1}%
}
\newcommand*{\colorboxedAux}[3]{%
  \begingroup
    \colorlet{cb@saved}{.}%
    \color#1{#2}%
    \boxed{%
      \color{cb@saved}%
      #3%
    }%
  \endgroup
}

\shorttitle{Surface forces in diffusiophoresis}
\shortauthor{S. Marbach, H. Yoshida and L. Bocquet}

\title{Surface forces on a diffusiophoretic particle}
\title{Local force and deformation on a phoretic particle}

\title{Local force balance and deformation of a particle during phoresis}
\title{Local and global force balance for diffusiophoretic transport}

\author{S. Marbach\aff{1}\aff{2},
  H. Yoshida\aff{3}
 \and L. Bocquet\aff{1} \corresp{\email{lyderic.bocquet@ens.fr}}}

\affiliation{\aff{1} Ecole Normale Sup\'{e}rieure, PSL Research University, CNRS, 24 rue Lhomond, Paris, France
\aff{2} Courant Institute of Mathematical Sciences, NYU, 251 Mercer Street, New York, NY, USA 
\aff{3} Toyota Central R$\&$D Labs., Inc., Bunkyo-ku, Tokyo 112-0004, Japan }

\begin{document}

\maketitle

\begin{abstract} % No more than 250 words
Electro- and diffusio- phoresis of particles correspond respectively to the transport of particles under electric field and solute concentration gradients. Such interfacial transport phenomena take their origin in a diffuse layer close to the particle surface, and the motion of the particle is force-free. 
In the case of electrophoresis, it is further expected that  the stress acting on the moving particle vanishes locally as a consequence of local electroneutrality. 
But the argument does not apply to diffusiophoresis, which takes its origin in solute concentration gradients. 
In this paper we investigate further the local and global force balance on a particle undergoing diffusiophoresis. We calculate the local tension applied on the particle surface %While it is known that the local force vanishes also for electrophoresis due to electroneutrality, we show here that, as a consequence of the osmotic origin of the driving forces, 
and show that, counter-intuitively, the local force on the particle does not vanish for diffusiophoresis, in spite of the global force being zero as expected. Incidentally, our description allows to clarify the osmotic balance in diffusiophoresis, which has been a source of debates in the recent years. We explore various cases, including hard and soft interactions, as well as porous particles, and provide analytic predictions for the local force balance in these various systems. 
The existence of local stresses  may induce deformation of soft particles undergoing diffusiophoresis, hence suggesting applications in terms of particle separation based on capillary diffusiophoresis.
%is a ubiquitous phenomenon. Most benchmark experiments are done on rigid particles such as colloids yet softer particles like polymers or DNA may also undergo phoretic motion. As these soft particles move, one could expect them to deform under hydrodynamic stresses. Yet for electrophoresis, it is widely known that the local force balance at the surface of the (soft) particle is ensured, and therefore, particles do not deform. However, here we show analytically that the local force at the surface of a particle undergoing Diffusiophoresis (in a gradient of solute) does not vanish. Therefore soft Diffusiophoretic particles should deform. We give analytic formulas for the local force on the surface in a number of cases, including for porous spheres representing polymers. This allows to gain physical intuition on the force balance and the relevant length scale where osmotic effects are at play. We expect these results to encourage innovative experiments to separate soft particles and to allow for better modeling of phoretic related effects.
\end{abstract}

%\begin{keywords}
%Authors should not enter keywords on the manuscript, as these must be chosen by the author during the online submission process and will then be added during the typesetting process (see http://journals.cambridge.org/data/\linebreak[3]relatedlink/jfm-\linebreak[3]keywords.pdf for the full list)
%\end{keywords}

%citep or citet

\section{Introduction}

\begin{figure}
\center
\includegraphics[width=0.9\textwidth]{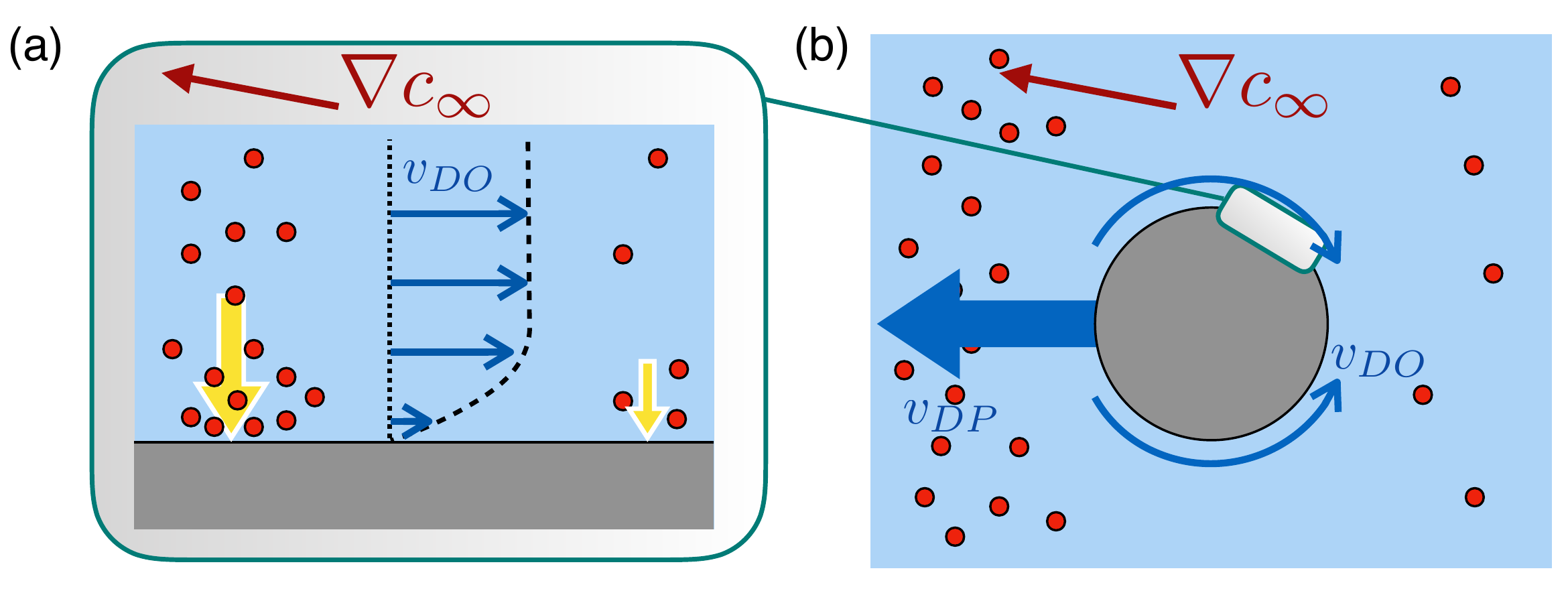}
\caption{\textbf{From diffusio-osmosis to diffusiophoresis}: (a) Schematic showing diffusio-osmotic flow generation. A surface (gray) is in contact with a gradient of solute (red particles). Here the particles absorb on the surface creating a pressure in the fluid (represented by yellow arrows). This pressure build-up is stronger where the concentration is highest, and induces a hydrodynamic flow $v_{DO}$ from the high concentration side to the low concentration side. (b) If this phenomenon occurs at the surface of a particle, the diffusio-osmotic flow will induce motion of the particle at a certain speed $v_{DP}$ in the opposite direction. This is called diffusiophoresis.}
\label{fig:OsmoToPhoresis}
\end{figure}

Phoresis corresponds to the motion of a particle induced by an external field, say $\Theta_\infty$: typically an electric potential for electrophoresis, a solute concentration gradient for diffusiophoresis, or a temperature gradient for thermophoresis~\citep{anderson1989colloid,ReviewOsmosis}. 
The particle velocity is accordingly proportional to the gradient of the applied field, writing in the general form
\begin{equation}
v_P= \mu_P \times (-\nabla \Theta_\infty)
\end{equation}
with $\Theta_\infty$ the applied field infinitely far from the particle. Phoretic motion has several key characteristics. First the motion takes its origin within the interfacial diffuse layer close to the particle: typically the electric double layer for charged particles, but any other surface interaction characterized by a diffuse interface of finite thickness.  Within this layer the fluid is displaced relatively to the particle due {\it e.g.} to electro-osmotic or diffusio-osmotic transport; see Fig.~\ref{fig:OsmoToPhoresis} for an illustration ~\citep{derjaguin1987some,anderson1989colloid}. Second, motion of the particle is force-free, {\it i.e.} the global force on the particle is zero, the particle moves at a steady velocity. 
This can be understood in simple terms for example for electrophoresis: the cloud of counter-ions around the particle experiences a force due to the electric field which is opposite to that applied directly to the particle, so that the total force acting on the system of the particle and its ionic diffuse layer experiences a vanishing total  
force. 
Both electro- and diffusio- phoresis and correspondingly electro- and diffusio-osmosis can all be interpreted as a single osmotic phenomena, since the two are related via a unique driving field, the electro-chemical potential \citep{ReviewOsmosis}. 
%\LB{revoir: Osmosis is a universal feature occurring in a broad number of domains ranging from biomedical applications to water supply challenges and with a number of daily-life implications~\citep{ReviewOsmosis}. %: from a number of biomedical challenges (kidney filtration, cell volume control, plant growth, desalination, drinking water) to a variety of artificial solutions (shell gas extraction, blue energy generation, laundry detergency) and so on
%It is traditionally explained as the force driving water through a membrane permeable to water only, when exposed to a gradient of concentration. Water flows towards the high concentration reservoir and equilibrates concentrations. One of the key aspects of osmosis is not really the membrane itself, but the fact that this membrane acts with a different strength on the solvent and the solute. Situations where differential forces act on the solvent and the solute occur especially at interfaces.} For example gradients of solutes in the presence of \textit{e.g.} an absorbing interface may induce flow motion in the bulk, this is called diffusio-osmosis -- see Fig.~\ref{fig:OsmoToPhoresis}-a. Symmetrically when a (solid) particle is suspended in a fluid, gradients of solute will induce motion of the particle -- see Fig.~\ref{fig:OsmoToPhoresis}-b. This motion is called "Diffusiophoresis"~\citep{derjaguin1987some,anderson1989colloid}. }

Interestingly these phenomena have gained renewed interest over the last two decades, in particular thanks to the development of microfluidic technologies which allow for an exquisite control of the physical conditions of the experiments, electric fields or concentration gradients. However, in contrast to electrophoresis, diffusiophoresis has been much less investigated since the pioneering work of Anderson and Prieve. Its amazing consequences in a broad variety of fields have only started to emerge, see \cite{ReviewOsmosis} for a review and ~\cite{abecassis2008boosting,palacci2010colloidal,palacci2012osmotic,velegol2016origins,moller2017steep,shin2018cleaning} as a few examples of applications. 
%At steady-state, diffusiophoresis is a force-free motion: the particle moves at constant velocity without any force acting on it, {\it i.e.} the global resulting force acting on the particle vanishes~\citep{anderson1989colloid}. 
The diffusiophoretic velocity of a particle under a (dilute) solute gradient writes 
\begin{equation}
\bm{v}_{DP} = \mu_{DP}  \times (-k_BT\bm{\nabla} c_\infty)
\label{eq:PhoresisDOneutral}
\end{equation}
where $ \mu_{DP} $ is the diffusiophoretic mobility, $\bm{\nabla} c_\infty$ is the solute gradient far from the sphere, $k_B$ Boltzmann's constant and $T$ temperature.
For example, for a solute interacting with a spherical particle via a potential $\mathcal{U}(z)$, where $z$ is the distance to the particle surface, the diffusiophoretic mobility writes~\citep{anderson1991diffusiophoresis}
\begin{equation}
\mu_{DP}=  - \frac{1}{\eta}\int_0^\infty  \, z \, \left(\exp\left(\frac{-\mathcal{U}(z)}{k_B T}\right) - 1\right) dz.
\label{KDP}
\end{equation}

In this work, we raise the question of the local and global force balance in phoretic phenomena, focusing in particular on diffusiophoresis. Indeed, while such interfacially-driven motions are force-free, {\it i.e.} the global force on the particle is zero, the local force balance is by no means obvious. For electrophoresis, it was discussed by \cite{long1996simultaneous} that local electroneutrality ensures  that the force acting on the particle is also vanishing locally in the case of a thin diffuse layer. 
Indeed the force acting on the particle is the sum of the electric force $dq\times\bm{E}_{\rm loc}$, with $dq$ the charge on an elementary surface and $\bm{E}_{\rm loc}$ the local electric field, and the hydrodynamic surface stress due to the electro-osmotic flow. To ensure mechanical balance within the electric double layer, this hydrodynamic stress has to be equal to the electric force on the double layer, which is exactly $- dq\,\bm{E}_{\rm loc}$ since the electric double layer carries an opposite charge to the surface. Therefore the local force on the particle surface vanishes. The absence of local force has some important consequences, among which we have the fact that particles such as polyelectrolytes undergoing electrophoresis do not deform under the action of the electric field~\citep{long1996simultaneous}. 

Such arguments do not obviously extend to diffusiophoresis. The main physical reason is that diffusiophoresis involves the balance of viscous shearing with an osmotic pressure gradient acting in the diffuse layer along the particle surface~ \citep{ReviewOsmosis}. 
While such a balance is simple and appealing, it led to various mis-interpretations and debates concerning osmotically-driven transport of particles~\citep{cordova2008osmotic,julicher2009comment,fischer2009comment,PhysRevLett.103.079802,PhysRevLett.102.159802,brady2011particle}, also in the context of phoretic self-propulsion  \citep{moran2017phoretic}. A naive interpretation of diffusiophoresis is that the particle velocity $v_{DP}$ results from the balance of Stokes' viscous force $F_v=6\pi\eta R v_{DP}$ and the osmotic force resulting from the osmotic pressure gradient integrated over the particle surface. The latter scales hypothetically as $F_{osm} \sim R^2\times R\nabla \Pi$, with $\Pi = k_BT c_\infty$ the osmotic pressure. Balancing the two forces, one predicts a phoretic velocity behaving as $v_{DP} \sim R^2 \frac{k_BT}{\eta} \nabla c_\infty$. Looking at the expression for the diffusiophoretic mobility in the thin layer limit, Eqs.~(\ref{eq:PhoresisDOneutral}) and~(\ref{KDP}), the latter argument does not match the previous estimate  by a factor of order $(R/\lambda)^2$, where $\lambda$ is the range of the potential of interaction between the solute and the particle. 
The reason why such a global force balance  argument  fails % is globally flawed 
is that  flows and interactions in interfacial transport occur typically over the thickness of the diffuse layer, in contradiction with the naive estimate above.  
%In his work in \citep{brady2011particle}, Brady tackled the question based on a "micromechanical" analysis of the solute and solvent transport in the presence of the colloidal particles. 

A second aspect which results from the previous argument is that the interplay between hydrodynamic stress and osmotic pressure gradient for diffusiophoresis may lead to a non-vanishing local surface force. Indeed in the absence of an electric force, only viscous shearing acts tangentially on the particle itself, while particle-solute neutral interactions are mostly acting on the orthogonal direction. A force tension may therefore be generated locally at the surface of the particle. This is in contrast to electrophoresis. %where $\UU$ is the potential of interaction between the solute and the particle, with typical range $\lambda$
%Interestingly, provided the value for the diffusio-osmotic mobility is constant over the particle's surface, it was shown by Morrison that this result holds for any particle shape (the argument is valid for any interfacially driven transport~\citep{morrison1970electrophoresis,anderson1989colloid}).
%This force-free motion has counter-intuitive implications and led to various mis-interpretations and debates concerning osmotically-driven transport of particles~\citep{cordova2008osmotic,julicher2009comment,fischer2009comment,PhysRevLett.103.079802,PhysRevLett.102.159802,brady2011particle}, in particular in the context of phoretic self-propulsion  \citep{moran2017phoretic}. 
% 

The question of global and local force balance in diffusiophoretic transport is therefore subtle and there is a need to clarify the mechanisms at stake. 
%properly solve the riddle. 
%To this end, one needs to go into the details of the force balance on the particle undergoing diffusiophoretic transport. 
% and identify the mechanical origin of the osmotic pressure.
In the derivations below we  first relax the hypothesis of a thin diffuse layer, and consider more explicitly the transport inside the diffuse layer, as was explored by various authors, using {\it e.g.} controlled asymptotic expansions \citep{sabass2012dynamics,sharifi2013diffusiophoretic,cordova2013osmotic}. 
Then on the basis of this general formulation, we are able to write properly the global and local force balance for diffusiophoresis.
Our results confirm the existence of a non-vanishing surface stress in diffusiophoresis, in spite of the global force being zero. 
%As we now discuss,  
%Here we obtain general analytic results for the hydrodynamic mobility and force balance {\it without assuming a thin diffuse layer}. We are able to rationalize the global force balance along the following lines. The difference between the two scalings originates %origin of this discrepancy actually takes its root 
%in the fact that for interfacially driven motion, the velocity gradient occurs mostly over a characteristic thickness $L_s$ close to the size of the diffuse layer $\lambda$, and not on  the particle size $R$, as {\it e.g.} for the Stokes flow. We report the expression for $L_s$ below. Looking into the details of the force balance we find that at the local scale equilibrium is not satisfied, meaning that the local force on a point of the surface is not zero. For a soft particle, this would result in deformation of the particle under diffusiophoresis. Furthermore, we 
To illustrate the underlying mechanisms, we consider a number of cases: diffusiophoresis under a gradient of neutral solutes, diffusiophoresis of a charged particle in an electrolyte bath, and diffusiophoresis of a porous particle. We also consider the situation of electrophoresis as a benchmark where the surface force on the particle is expected to vanish. We summarize our results in the next section and report the detailed calculations in the sections hereafter.

\section{Geometry of the problem and main results: surface forces on a phoretic particle}

\subsection{Diffusiophoretic velocity}

We consider a sphere of radius $R$ in a solution containing one or multiple solutes, charged or not. The surface of the sphere interacts with the species over a typical lengthscale $\lambda$, via \textit{e.g.} electric interactions, steric repulsion or any other interaction. In the case of diffusiophoresis, a gradient of solute, $\nabla c_\infty$, is established at infinity along the direction $z$. The sphere moves accordingly at constant velocity $v_{DP} \bm{e}_z$ and we place ourselves in the sphere's frame of reference. 
%\LB{extending on similar work performed for electrophoresis \citep{ohshima1983approximate}}.
% was done for electrophoresis in the case of 
%\LB{citer Ohshima et al. \citep{ohshima1983approximate}: electrophoresis, weak potential but no thin EDL}
%
We consider that the interaction between the solute and the particle occurs via a potential ${\cal U}$, so that Stokes' equation for the fluid surrounding the sphere writes
 \begin{equation}
\eta \bm{\nabla^2} \bm{v} - \bm{\nabla} p +c(\bm{r}) (-\bm{\nabla} {\cal U})=0.
\label{Stokesplus}
\end{equation}
The boundary conditions {\it on the particle's surface} are the no-slip boundary condition  (note that the no-slip boundary condition may be relaxed to account for partial slip at the surface, in line with~\cite{ajdari2006giant}), complemented by the prescribed velocity at infinity (in the frame of reference of the particle):
%\begin{eqnarray}
%&\mathbf{v} (r=R)=0, \nonumber \\
%&\mathbf{v}(r\rightarrow \infty)= -\bm{v}_{DP}
%\end{eqnarray}
\begin{equation}
\bm{v} (r=R)= \bm{0} \,\,\;\mathrm{and}\,\,\; \bm{v}(r\rightarrow \infty)= -\bm{v}_{DP}
\end{equation}
The solute concentration profile obeys a Smoluchowski equation in the presence of the external potential ${\cal U}$, in the form
\begin{equation}
0=-\bm{\nabla}\cdot \left[-D_s\bm{\nabla} c + \frac{D_s}{k_B T} c (-\bm{\nabla} {\cal U})\right]
\end{equation}
where $D_s$ is the diffusion coefficient of the solute, with the boundary condition at infinity accounting for a constant solute gradient $c(r\rightarrow \infty)\simeq c_0 + r\cos\theta \nabla c_\infty$; $c_0$ is a reference concentration. Note that we neglected convective transport here, assuming a low P\'eclet regime. In this case, the Smoluchowski equation is self-consistent and provides a solution for the solute concentration field, which therefore acts as an independent source term for the fluid equation of motion in Eq.~\eqref{Stokesplus}. 

\begin{figure}
\center
\includegraphics[width=0.9\textwidth]{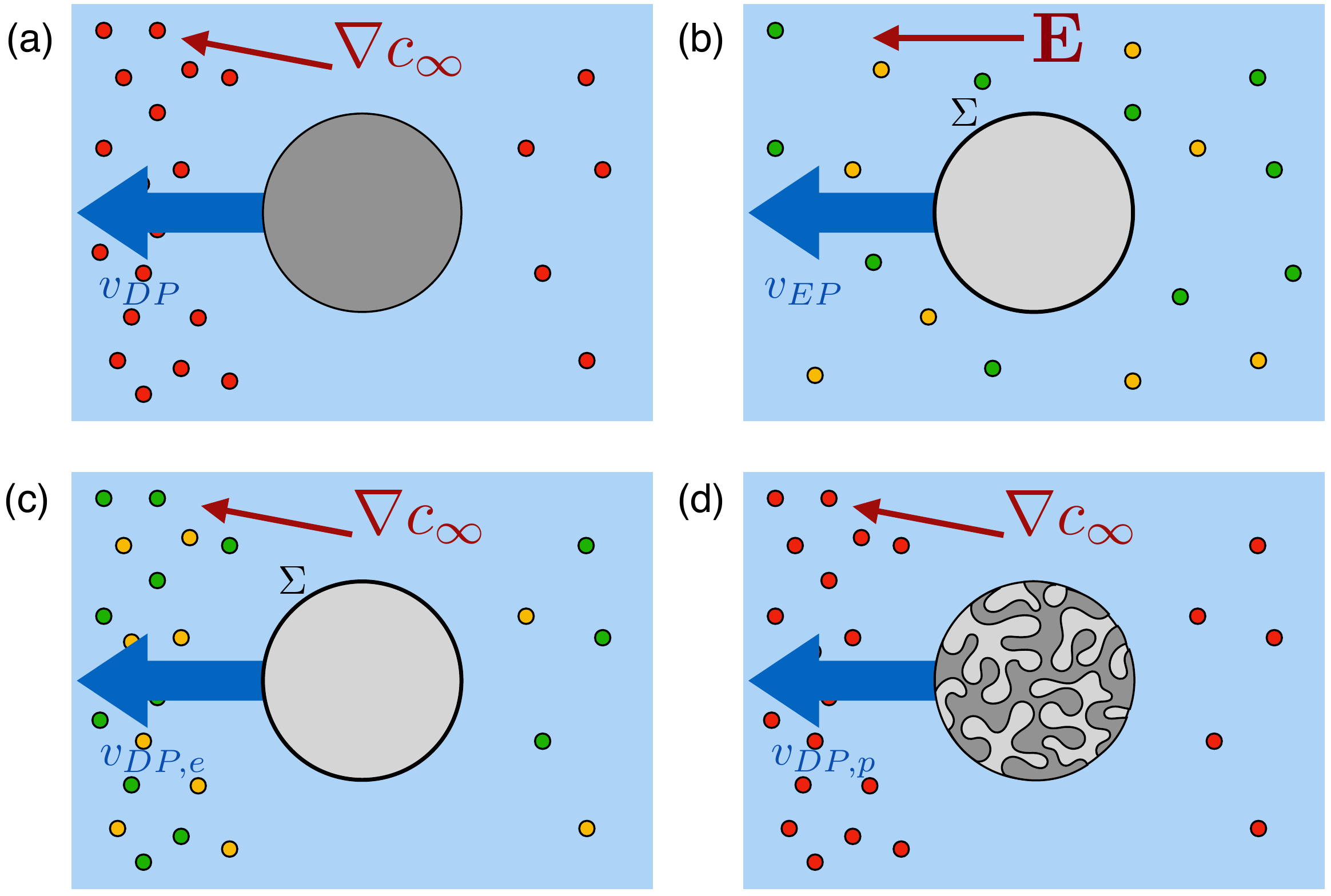}
\caption{\textbf{Geometries considered in this paper}: (a)  diffusiophoresis under neutral solute gradients: a spherical particle moving in a (uncharged) solute gradient. (b) electrophoresis: a spherical particle with surface charge $\Sigma$ moving in an electric field in a uniform electrolyte. (c)  diffusiophoresis under ionic concentration gradients : a spherical particle with surface charge $\Sigma$ moving in an electrolyte gradient. (d) diffusiophoresis of a porous particle: a porous spherical particle moving in an uncharged solute gradient.}
\label{fig:AllCases}
\end{figure}

In this paper we report analytic results in various cases as represented in Fig.~\ref{fig:AllCases}. First (see Fig.~\ref{fig:AllCases}-a), we show that for any radially symmetric potential ${\cal U}(r)$, one may compute an exact solution of \eqref{Stokesplus} for the velocity profile and the local force. Second, going to more general electro-chemical drivings, like electrophoresis (see Fig.~\ref{fig:AllCases}-b) or diffusiophoresis of a charged sphere in an electrolyte solution (see Fig.~\ref{fig:AllCases}-c), it is also possible to compute exact solutions, assuming a weak driving force with respect to equilibrium. Finally, we come back to simple diffusiophoresis of a porous sphere with a radially symmetric potential ${\cal U}(r)$ (see Fig.~\ref{fig:AllCases}-d) and give similar analytic results. The porosity of the sphere is accounted for by allowing flow inside the sphere with a given permeability. %In that case it also possible to derive exact solutions with no further assumption. 

\subsection{Phoretic velocity}

We summarize briefly the analytic results for the phoretic velocity in the various cases considered. Results are reported in Table~\ref{TableVelocity}. 

\vspace{2mm}

\noindent \textit{Diffusiophoresis under gradients of a neutral solute}. For any radially symmetric potential ${\cal U}(r)$, one may compute an exact solution of \eqref{Stokesplus} for the velocity profile by extending textbook techniques for the Stokes problem in \cite{happel2012low} (see also \cite{ohshima1983approximate} for a related calculation in the context of electrophoresis). It can be demonstrated that the solution for $\bm{v}(\bm{r})$ involves a Stokeslet as a leading term, which allows to calculate the force along the axis of the gradient as the prefactor of the Stokeslet term ($v\sim F/r$). This allows to deduce the global force on the particle as
\begin{equation}
F= 6 \pi R \eta v_{DP} 
-  2 \pi R^2  \int_R^{\infty} c_0(r) (-\partial_r{\cal U})(r) \times \varphi(r) dr  
\label{FDO}
\end{equation}
with $\varphi(r)=\frac{r}{R} - \frac{R}{3r} - \frac{2}{3} \left(\frac{r}{R}\right)^2 $ a dimensionless function, the factor $\frac{2}{3}$ originating from the angular average, and the function $c_0(r)$ is such that the concentration profile writes $c(r,\theta)=c_0+c_0(r)\,{\cos \theta}$. Eq.~(\ref{FDO}) decomposes as the sum of the classic Stokes friction force on the sphere and a balancing force of osmotic origin, taking its root in the  interaction $\mathcal{U}$ of the solute with the particle. The steady-state diffusiophoretic velocity results from the force-free condition, $F=0$, and therefore writes
%If we require that the total force on the sphere vanish, $F_z = 0$ then we have
\begin{equation}
v_{DP} = \frac{2 \pi R^2}{6\pi\eta R}   \int_R^{\infty} c_0(r) (-\partial_r{\cal U})(r) \times \varphi(r) dr 
\label{VDPexact}
\end{equation} 
Remembering that $c_0(r) \propto R\nabla c_\infty$, this equation generalizes Eq.~(\ref{eq:PhoresisDOneutral}) obtained in the thin layer limit. Note that \eqref{VDPexact} is very similar to Eq.~(2.7) in \cite{brady2011particle}, with the $r$-dependent term $2 \pi R^2\times \varphi(r)$ replaced in \cite{brady2011particle} by the prefactor $L(R)$. 
However the integrated "osmotic push" is weighted here by the local factor $\varphi(r)$ (in contrast to \cite{brady2011particle}) and this detail actually changes the whole scaling for the mobility.

\begin{center}
\begin{tabular}{|c|c|}
\hline 
\textbf{Diffusiophoresis of colloids } & $\displaystyle v_{DP} = \frac{R}{3\eta} \int_R^{\infty} \frac{c(r,\theta) - c_0}{\cos\theta} (-\partial_r \mathcal{U})  \,\varphi(r) dr$ \\
\textit{neutral solutes} &  \\
\textit{with soft interaction potential} $\mathcal{U}(r)$ & $\mathrm{with} \,\, \varphi(r) =  \frac{r}{R} - \frac{R}{3r} - \frac{2r^2}{3 R^2} $ \\ 
\hline 
\textit{with the thin layer approximation$^{\star}$} & 
%$v_{DP,\lambda} = \displaystyle \nabla c_{\infty} \frac{k_B T}{\eta } \int_0^{\lambda} \left( e^{-\beta \mathcal{U}(z)} - 1 \right) z dz $ \\
$v_{DP,\lambda} = \displaystyle \nabla c_{\infty} \frac{k_B T}{\eta } \int_0^{\infty} \left( e^{-\beta \mathcal{U}(z)} - 1 \right) z dz $ \\
\hline
\textbf{Generalized formulation} & $\displaystyle v_{P} = \frac{R}{3\eta} \int_R^{\infty}  \left(\sum_{\mathrm{species} \, i} \partial_r \rho_{0,i}\times \tilde{\mu}_i  \right) \,\, \varphi(r)dr$ \\
\textit{} &  \\
\textit{see Sec. 4} & $\mathrm{with} \,\, \varphi(r) =  \frac{r}{R} - \frac{R}{3r} - \frac{2r^2}{3 R^2} $ \\ 
\hline 
\textbf{Diffusiophoresis of porous colloid}  & $\displaystyle v_{DP,p} = \frac{R}{3\eta} \int_0^{\infty} \frac{c(r,\theta) - c_0}{\cos\theta} (-\partial_r\mathcal{U})  \,\, \Phi(r) dr$  \\ 
\textit{ neutral solutes} &  \\
\textit{with soft interaction potential $\mathcal{U}(r)$} & $\displaystyle \mathrm{with} \,\, \Phi(r)$ defined in Eq.~(\ref{psiPhoresis}) \\
%\hline 
%\textit{Limit of low permeance $\kappa$} & $\displaystyle v_{DP,p} \geq v_{DP}$ \\ 
\hline
\end{tabular} 
\captionof{table}{Main results for the phoretic velocity of plain and porous colloidal particles. Here $\tilde{\mu}_i$ is field which is the perturbation to the chemical potential of species $i$ under the applied field, {\it i.e.} $\tilde{\mu}_i\propto \nabla \mu_\infty$ the applied electro-chemical gradient at infinity; $\rho_{0,i}$ is the concentration profile of specie $i$ in equilibrium.  $^{\star}$Note that this result is similar to the diffusio-osmotic velocity over a plane surface reported in \cite{anderson1991diffusiophoresis}.}
\label{TableVelocity}
\end{center}

\vspace{2mm}

\noindent \textit{Generalized formula for phoresis under electro-chemical gradients}. 
It is possible to generalize the previous results to charged species under an electro-chemical potential gradient. 
The general expression for the diffusiophoretic velocity is written in terms of  the electro-chemical potential $\mu_i $ (where $i$ stands for each solute specie $i$). 
One may separate the electro-chemical potential  as $\mu_i = \mu_{0,i} + \tilde{\mu}_i$, where $\mu_{0,i}$ is the equilibrium chemical potential and $ \tilde{\mu}_i$ the perturbation due to an external field, so that 
$ \tilde{\mu}_i \propto \nabla \mu_\infty$, the applied electro-chemical potential gradient at infinity. The derivation assumes a weak perturbation,
% one contained in the term $ \tilde{\mu}_i$, is weak as compared to the equilibrium chemical potential, so that 
 $ \tilde{\mu}_i \ll \mu_{0,i}$. 
% it is possible to generalize the previous results and obtain a  general expression for the phoretic velocity,
This leads to an expression of the generalized expression for the diffusiophoretic velocity in a compact form
\begin{equation}
\displaystyle v_{P} = \frac{R}{3\eta} \int_R^{\infty}  \left(\sum_{\mathrm{species} \, i} \partial_r \rho_{0,i} \times \tilde{\mu}_i  \right) \,\, \varphi(r)dr
\end{equation} 
where $\rho_{0,i}$ is the concentration profile at equilibrium.
Details of the calculations are reported in Sec.~4.

\vspace{2mm}

\noindent \textit{Diffusiophoresis of a porous sphere}. It is possible to extend the derivation to the case of a porous colloid. This may be considered as a coarse-grained model for a polymer. We assume in this case
that the solute is neutral and interacts with the sphere via a radially symmetric potential $\mathcal{U}$. In that case the Stokes equation~(\ref{Stokesplus}) is extended inside the porous sphere with the addition of a Darcy term:
  \begin{equation}
\eta \bm{\nabla^2} \bm{v} - \frac{\eta}{\kappa} \bm{v}- \bm{\nabla} p +c(\bm{r}) (-\bm{\nabla} {\cal U})=0
\label{Stokesplusporous}
\end{equation}
where $\kappa$, expressed in units of a length squared, is the permeability of the sphere. 
The expression for the diffusiophoretic velocity can be calculated explicitly, with an expression formally similar to the diffusiophoretic velocity, 
\begin{equation}
v_{DP,p} = \frac{R}{3\eta} \int_0^{\infty} \frac{c(r,\theta) - c_0}{\cos\theta} (-\partial_r\mathcal{U})  \,\, \Phi(r) dr
\end{equation}
where the details of the porous nature of the colloid are accounted for in the weight $\Phi(r)$, as reported in Eq.~(\ref{psiPhoresis}). The latter is a complex function of $k_\kappa R$, where $k_\kappa=1/\sqrt{\kappa}$ is the inverse screening length associated with the permeability of the colloid, with
radius $R$. 
Details of the calculations are reported in Sec.~5.

%We find that a low permeances the diffusiophoretic velocity is always increased as compared to the fully impermeable sphere. At high permeances, the diffusiophoretic velocity may reverse (depending on the details of the interaction potential). This result points therefore to a wealth of potential behaviors for diffusiophoresis of a porous/soft particle.

\subsection{Local force balance on the surface}

Beyond the diffusiophoretic velocity, the theoretical framework also allows to compute the global and local forces on the particle. 
%\noindent \textit{Local force balance broken at the particle's surface}
Writing the local force balance at the particle surface, we find in general that the particle withstands a local force that does not  vanish for diffusiophoresis. The local force $d {\bm f}$ on an element of surface $dS$ of a phoretic particle can be written generally as
\begin{equation}
d {\bm f} = \left( - p_0 + \frac{2}{3} \pi_s \cos \theta \right) dS \, {\bm e_r}  + \left( \frac{1}{3} \pi_s \sin \theta \right)  dS  \, {\bm e_{\theta}} 
\label{eq:localforce1}
\end{equation}
where the local force is fully characterized by a force per unit area - or pressure - $\pi_s$. In this expression $p_0$ is the bulk hydrostatic pressure and ${\bm e_r}$ and $ {\bm e_{\theta}}$ are the unit vectors in the spherical coordinate system centered on the sphere. We report  the value of $\pi_s$ in the table below for the various cases considered, see Table~\ref{TableFs}. While the surface force is found to be non-vanishing for all diffusiophoretic transport, our calculations show that $\pi_s\equiv 0$  for electrophoretic driving: a local force balance is predicted for electrophoresis 
%where local force balance is ensured. This was first discussed in 
in agreement with the argument of in \cite{long1996simultaneous} (see the details in Sec. 4). 

Let us report more specifically the  results for the local force in the different cqses.
\vspace{2mm}

\noindent \textit{Local force for diffusiophoresis with neutral solutes -- }
For solutes interacting with the colloid via a soft interaction potential $\mathcal{U}(r)$, one finds that the  surface force takes the
form
\begin{equation}
%\pi_s =  \int_R^{R+\lambda} c_0(r) (-\partial_r{\cal U})(r) \psi(r) dr
\pi_s =  \int_R^{\infty} c_0(r) (-\partial_r{\cal U})(r) \psi(r) dr
\label{eq:fsDP}
\end{equation}
where $\psi(r) =   \frac{R}{r} - \frac{r^2}{R^2} $ is a geometrical factor. 
As we demonstrate in the following sections, in the case of a thin double layer, the local force reduces to a simple and transparent expression:
%In the thin diffuse layer limit, with $r-R \sim\lambda \ll R$, one may expand $\psi(r) \simeq -2 (r-R)^2/R^2$, while the concentration profile $c_0(r)$ can be approximated as 
%\begin{equation}
%c_0(r)\simeq R \nabla c_\infty \times \left[\frac{r}{R} + \frac{1}{2} \left(\frac{R}{r}\right)^2\right] \exp[-{\cal U}(r)/k_BT].
%\end{equation}
%The weight $\psi(r)\sim \lambda^2/R^2$ is a signature of the fact that the velocity gradient occurs on the width $\lambda$ and not on the particle size $R$. An osmotic pressure is indeed expressed at the particle's surface and yields diffusiophoretic transport, but in a very subtle way which does not reduce to considering only the direct solute force.
% %to the subtle force balance underlying the diffusiophoretic transport, which 
% This corrects the naive argument suggested in the introduction. The local force is in fact the sum of the hydrodynamic shear force, normal pressure and direct interaction with the solute. Using the exact results for the velocity profile in  the thin layer regime, $\lambda\ll R$, one finds
%%Now expanding and simplifying yields the force over the solid angle
\begin{equation}
\pi_s \simeq  \frac{9}{2}   k_B T L_s  \nabla c_{\infty}
\label{eqForceLocale}
\end{equation}
where 
$L_s = \int_R^{\infty} \left( e^{-\beta \mathcal{U}(z)} - 1\right) dz$ has the dimension of a length and quantifies the excess adsorption on the interface. %Eq.~\eqref{eqForceLocale} can be recovered easily with a simplistic argument: one expects this osmotic force to scale as $\mathcal{V}_{\rm int} \nabla \Pi = \mathcal{V}_{\rm int}  \nabla (k_B T c_{\infty})$ where $\mathcal{V}_{\rm int} $ is the interaction volume. Writing $L_s$ the typical interaction lengthscale we have $\mathcal{V}_{\rm int} \approx 4\pi R^2 L_s$, leading accordingly to Eq.~\eqref{eqForceLocale} since $dS \simeq R^2$. 

%%Note that $\Delta \Pi = L_s k_B T \nabla c_{\infty}$
\vspace{2mm}

\noindent \textit{Local force for phoresis under small electro-chemical gradients --}
As for the velocity, it is possible to generalize the previous results to the case of a general, small, electro-chemical driving. 
In the case of a thin diffuse layer, the result for $\pi_s$ takes the generic form
\begin{equation}
\displaystyle \pi_s =  \int_R^{\infty}\left(\sum_{\mathrm{species} \, i} \partial_r \rho_{0,i}\times \tilde{\mu}_i  \right)\,\, \psi(r)  dr
\end{equation}
with $\psi(r) =  \frac{R}{r} - \frac{r^2}{R^2} $ and we recall that $\tilde{\mu}_i \propto \nabla\mu_\infty$ the gradient of the electro-chemical potential far from the colloid. This result applies to both diffusio- and electro- phoresis. As reported in the Table~\ref{TableFs}, the local force is non-vanishing for diffusiophoresis but for electrophoresis one predicts $\pi_s\equiv 0$. 
%It is interesting for instance to comment the value obtained in the case of diffusiophoresis of a charged particle in a gradient of electrolyte. In that case the length scale of the interaction is characterized by the Debye length $\lambda_D$ ( writing $\lambda_D^{-2} =  \frac{e^2 c_0}{\epsilon k_B T}$). In the thin layer regime $\lambda_D \ll R$ one finds as expected that the relevant length of the osmotic push is $\lambda_D$ since  
%\begin{equation}
%\pi_s \simeq  \frac{9  \, Du^2}{4} k_B T\lambda_D \nabla c_{\infty} .
%\end{equation}
%Notably the local force is modulated by the non-dimensional factor $Du = \Sigma/e \lambda_D c_0$, where $\Sigma$ is the surface charge of the particle. Gathering all contributions in concentration gives a scaling of $\pi_s \propto \nabla \left( 1/\sqrt{c}\right)$ which is rather surprising. Note that in the small debye length regime we expect $c$ to be rather large. This non-linearity reveals that deformation of the particles is "power-sensing" and therefore dependent on the absolute position of the particle. One may therefore expect similar spectacular behaviors as for diffusiophoretic "log-sensing"~\citep{palacci2012osmotic,shin2018cleaning}.

\begin{center}
\begin{tabular}{|c|c|}
\hline 
\textbf{Diffusiophoresis of colloids}  & $\displaystyle \pi_s =  \int_R^{+\infty} \frac{c(r,\theta) - c_0}{\cos\theta} \partial_r(-\mathcal{U})\,\, \psi(r) dr$ \\
\textit{neutral solutes} &  \\
\textit{with soft interaction potential} $\mathcal{U}(r)$ & $\mathrm{with} \,\, \psi(r) =  \frac{R}{r} - \frac{r^2}{R^2} $ \\ 
\hline 
\textit{ thin layer approximation} & 
$\displaystyle \pi_s = \frac{9}{2} L_s k_B T \nabla c_{\infty} $ \\
&\, $\mathrm{with} \,\, L_s = \int_0^{\infty} \left( e^{-\beta \mathcal{U}(z)} - 1\right) dz$ \\
\hline
\textbf{Generalized formulation} & $\displaystyle \pi_s =  \int_R^{\infty}\left(\sum_{\mathrm{species} \, i} \partial_r \rho_{0,i}\times \tilde{\mu}_i  \right)\,\, \psi(r)  dr$ \\
\textit{ thin layer approximation}&  \\
\textit{see Sec. 4} & $\mathrm{with} \,\, \psi(r) =  \frac{R}{r} - \frac{r^2}{R^2} $ \\ 
\hline 
\textit{Electrophoresis} & $\pi_s = 0$\\ 
\hline 
\textit{Diffusiophoresis} & $\pi_s =  \frac{9 \, Du^2}{4} k_B T \lambda_D \nabla c_{\infty}  $ \\
\textit{charged colloid with surface charge $\Sigma$} &   \\
%$\pi_s(\lambda_D \gg R) \simeq \frac{\, Du^2}{2} k_B T \frac{R^2}{\lambda_D} \nabla c_{\infty} $
 \textit{(thin Debye layer limit)} & $Du = \Sigma/e \lambda_D c_0$ \\
\hline 
\textbf{Diffusiophoresis of a porous colloid} & $\displaystyle \pi_s =  \int_R^{+\infty} \frac{c(r,\theta) - c_0}{\cos\theta} \partial_r(-\mathcal{U})\,\, \Psi(r) dr$ \\
\textit{neutral solutes} &  \\
\textit{with soft interaction potential} $\mathcal{U}(r)$  & $\displaystyle \mathrm{with} \,\, \Psi(r)$  defined in Eq.~(\ref{forcePsi})\\
\hline 
%\textit{Limit of low permeance $\kappa$} & $\displaystyle \pi_s(\kappa \rightarrow 0) \leq \pi_s(\kappa = 0)$ \\ 
%\hline 
\end{tabular} 
\captionof{table}{Main results for the  local  surface force on plain and porous colloidal particles undergoing phoretic transport. Here $\tilde{\mu}_i$ is the perturbation to the chemical potential of species $i$ and $\rho_{0,i}$ its concentration profile in equilibrium. Note that $\lambda_D$ is the Debye length ($\lambda_D^{-2} =  \frac{e^2 c_0}{\epsilon k_B T}$) and $Du = \Sigma/e \lambda_D c_0$ is a Dukhin number.}
\label{TableFs}
\end{center}

\vspace{2mm}

\noindent \textit{Local force for diffusiophoresis of a porous particle --}
Finally for a porous colloid undergoing diffusiophoresis, the local force is a function of the permeability and the diffusion coefficient of the solute inside and outside the colloid, say $D_1$ and $D_2$. The general formula writes as
\begin{equation}
 \pi_s =  \int_R^{+\infty} \frac{c(r,\theta) - c_0}{\cos\theta} \partial_r(-\mathcal{U})\,\, \Psi(r) dr
\end{equation}
where the expression for the function $\Psi(r)$ is given in Eq.~(\ref{forcePsi}). This is a  quite cumbersome expression in general, 
but in the thin diffuse layer limit, and small permeability $\kappa$ of the colloid, the local force takes a simple form
\begin{equation}
\pi_s(\kappa \rightarrow 0) =  \pi_s(\kappa = 0) \times \frac{D_2}{D_2 + D_1/2} \left(1 - \frac{2}{k_{\kappa} R} \right)
\end{equation}
where $  \pi_s(\kappa = 0) = \frac{9}{2} L_s k_B T \nabla c_{\infty}$; $k_{\kappa}=1/\sqrt{\kappa}$ is the inverse screening length associated with the Darcy flow inside the porous colloid, and $R$ is the particle radius. 
%This shows that the local force decreases with porosity. 
%In fact, the increase of the diffusiophoretic velocity with porosity calls for a decrease of the local force or friction. 

\vspace{2mm}

{In the next sections we detail the calculations leading to  the results in Tables \ref{TableVelocity} and~\ref{TableFs}. 
}

\section{Diffusiophoresis of a colloid under a gradient of neutral solute}
\label{sec:DP}

\begin{figure}[b!]
\centering
 \includegraphics[width=0.6\textwidth]{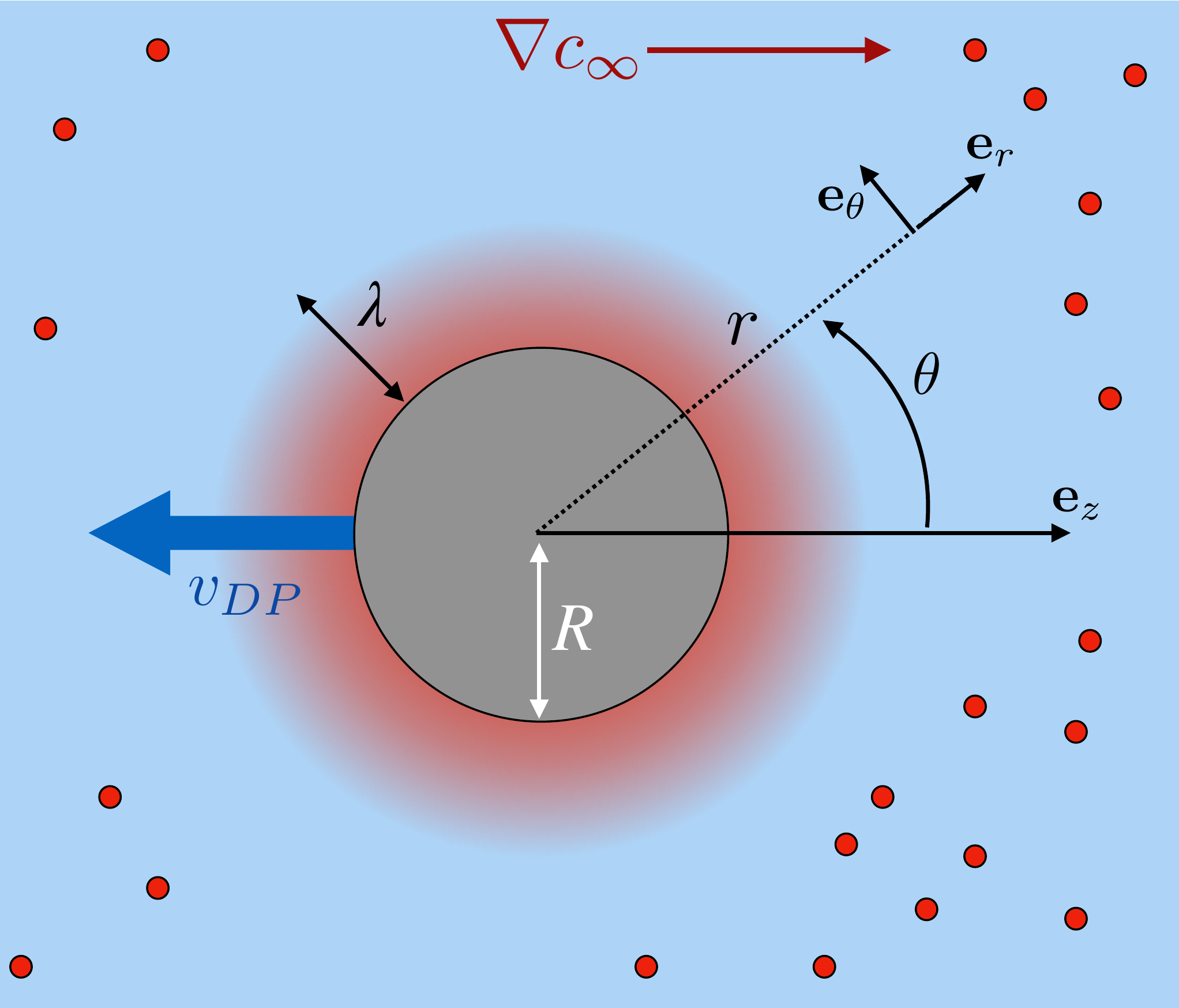}
 \caption{\textbf{Schematic of the coordinate system for the diffusiophoretic sphere.} The sphere interacts with the solute via a potential $\mathcal{U}(r)$ over a range $\lambda$. }
 \label{fig:schem}
\end{figure}

We focus first on diffusiophoresis of an impermeable particle, see Fig.~\ref{fig:AllCases}-a, under a concentration gradient of neutral solute. 
The solute  interacts with the particle via a soft interaction potential $\mathcal{U}(r)$ which only depends on the radial coordinate $r$ (with the origin  at the sphere center). In order to simplify the calculations we will consider that the interaction potential is non-zero only over a finite range, from the surface of the sphere $r=R$ to some boundary layer $r=R + \lambda$: the range $\lambda$ is finite but not necessarily small as compared to $R$, see Fig.~\ref{fig:schem}. 
One may take $\lambda\rightarrow \infty$ at the end of the calculation.

In the far field, the solute concentration obeys $\bm{\nabla} c |_{r \rightarrow \infty} = \nabla c_{\infty} \bm{e_z}$. The geometry is  axisymmetric and in  spherical coordinates, one may write $c(r \rightarrow \infty,\theta) = c_0 + \nabla c_{\infty} r\, \cos \theta$. Considering the boundary conditions for the concentration and the symmetry of the potential $\mathcal{U}$, one expects that the concentration can be written as $c(r,\theta) = c_0 + R\nabla c_{\infty}\times  f(r) \cos \theta$ where $f(r)$ is a radial and dimensionless function, which remains to be calculated. 

Note that in the following we neglect convection of the solute within the interfacial region, which may modify the steady-state concentration field of the solute around the particle. However such an assumption is generally valid because the P\'{e}clet number built on the diffuse layer is expected to be small. Our results could however be extended to include this effect on the mobility as a function of a (properly defined) P\'{e}clet number, as introduced in \cite{anderson1991diffusiophoresis,ajdari2006giant,sabass2012dynamics,michelin2014phoretic}. Similarly the effect of hydrodynamic fluid slippage at the particle surface may be taken into account, in line with the description in \citep{ajdari2006giant}.

\subsection{Flow profile}

\subsubsection{Constitutive equations for the flow profile}

The flow profile around the sphere is incompressible $\mathrm{div} (\vec{v}) = 0$ and obeys Stoke's equation, Eq.~(\ref{Stokesplus}). The projection of the Stokes equation along the unit vectors $\bm{e}_r$ and $\bm{e}_{\theta}$ gives 
\begin{equation}
\begin{cases}
\displaystyle \eta\left( \Delta v_r - \frac{2v_r}{r^2} - 2\frac{v_{\theta} \cos \theta}{r^2 \sin \theta} - \frac{2}{r^2} \frac{\partial v_{\theta}}{\partial \theta} \right) = \partial_r p - c(r,\theta) \partial_r(-\mathcal{U}) \\
\displaystyle \eta \left( \Delta v_{\theta} - \frac{v_{\theta}}{r^2\sin^2\theta} + \frac{2}{r^2}\frac{\partial v_r}{\partial \theta}\right) = \frac{1}{r}\partial_{\theta} p 
\end{cases}
\label{eq:StokesDP}
\end{equation}
The boundary conditions for the flow are (i) the prescribed diffusiophoretic flow far from the sphere, and (ii)  impermeability and  no slip condition on the particle surface:
\begin{equation}
\begin{cases}
v_r(r \rightarrow \infty,\theta) =  - v_{DP} \cos \theta \,\,\, \mathrm{and} \,\,\,  v_{\theta}(r \rightarrow \infty,\theta) =   v_{DP} \sin \theta \\
v_r(r = R,\theta) = 0  \,\,\, \mathrm{and} \,\,\, v_{\theta}(r = R,\theta) = 0 
\end{cases}
\end{equation}

\subsubsection{Solution for the flow profile}

We define a potential field $\psi$ such that
\begin{equation}
v_r = \frac{1}{r^2\sin\theta} \partial_{\theta} \psi \,\,\, \mathrm{and} \,\, v_{\theta} = - \frac{1}{r \sin \theta} \partial_r \psi
\end{equation}
so that the incompressibility condition $\mathrm{div} (\vec{v}) = 0$ is accordingly verified. 
We can rewrite the Stokes equations using the operator $E^2 = \frac{\partial^2}{\partial r^2} + \frac{\sin \theta}{r^2} \frac{\partial}{\partial \theta} \left( \frac{1}{\sin \theta} \frac{\partial}{\partial \theta} \right)$ as
\begin{equation}
\begin{cases}
\displaystyle \frac{\eta}{r^2 \sin \theta} \partial_{\theta} \left[ E^2 \psi \right] = \partial_r p - c(r,\theta) \partial_r(-\mathcal{U}) \\
\displaystyle  \frac{- \eta}{r \sin \theta} \partial_r   \left[ E^2 \psi \right] = \frac{1}{r}\partial_{\theta} p .
\end{cases}
\label{eq:4flow2}
\end{equation}
Adding up derivatives of the above formula allows to cancel the pressure contribution and obtain the simple equation for the potential field
\begin{equation}
\eta E^4 \psi = - \sin \theta \frac{\partial c(r,\theta)}{\partial \theta} \partial_r(-\mathcal{U})
\label{eq:4flow3}
\end{equation}
Using the general expression for $c(r,\theta)$, one obtains
\begin{equation}
\eta E^4 \psi =  \sin^2 \theta R\nabla c_{\infty} \, f(r) \partial_r(-\mathcal{U})
\end{equation}
We may therefore look for $\psi$ as $\psi = F(r) \sin^2\theta$ and we note that $E^2 \psi = \tilde{E}^2 F(r) \sin^2\theta$ where $\tilde{E}^2 = \frac{\partial^2}{\partial r^2} -  \frac{2}{r^2}$ so that
\begin{equation}
\tilde{E}^4 F(r) =  \frac{\nabla c_{\infty} R}{\eta} f(r) \partial_r(-\mathcal{U})
\end{equation}
We introduce $\tilde{f}(r) =  \frac{\nabla c_{\infty} R}{\eta} f(r) \partial_r(-\mathcal{U})$. Like the potential $\mathcal{U}(r)$, $\tilde{f}(r)$ is a compact function that is non-zero  only over the interval $[R;R + \lambda]$. 
The general solution of this equation is
\begin{equation}
\begin{split}
F(r) = &\frac{A}{r} + B r + r^2 C + D r^4 - \frac{1}{r}\int_R^r \frac{\tilde{f}(x)x^4}{30}dx + r \int_R^r \frac{\tilde{f}(x)x^2}{6}dx \\
&- r^2 \int_R^r \frac{\tilde{f}(x)x}{6}dx + r^4 \int_R^r \frac{\tilde{f}(x)}{30x}dx 
\end{split}
\end{equation}
where $A, B, C$ and $D$ are integration constants to be determined by the boundary conditions.  Note that the integrals do not diverge since $\tilde{f}$ is defined on a compact interval. The condition that the flow has to be finite far from the sphere $r\rightarrow \infty$ yields immediately
\begin{equation}
D = - \int_R^{R+\lambda} \frac{\tilde{f}(x)}{30x}dx \,\,\, \mathrm{and} \,\,\, C =  \int_R^{R+\lambda} \frac{\tilde{f}(x)x}{6}dx - {v_{DP}\over 2}
\end{equation}
Impermeability and no slip boundary conditions are equivalent to  $F(R) = F'(R) = 0$. This gives the values of $A$ and $B$ and the flow is now fully specified as 
\begin{equation}
\begin{split}
F(r) = &\frac{A}{r} + B r - \frac{ r^2}{2} v_{DO} + \int_R^r \tilde{f}(x) \left(  \frac{ r x^2}{6} - \frac{x^4}{30r} \right) dx +  \int_{R+\lambda}^r \tilde{f}(x)\left( \frac{r^4}{30x} - r^2\frac{x}{6}  \right) dx  \\
\mathrm{with} \, \, & A = - \frac{R^3}{4} v_{DO} +  \int_R^{R+\lambda} \tilde{f}(x) \left( \frac{R^3x}{12} - \frac{R^5}{20 x}  \right) dx  \\
\mathrm{and} \, \, & B = \frac{3 R}{4} v_{DO} +  \int_R^{R+\lambda} \tilde{f}(x) \left( \frac{R^3}{12x} - \frac{R x}{4} \right) dx 
\end{split}
\end{equation}

This provides an explicit expression for 
the flow profile as %can then be fully expressed simply (and in a similar way to the canonical Stokes problem of friction on a sphere) 
\begin{equation}
\begin{cases}
v_r &= \sin \theta \left( 2 \frac{\tilde{B}(r)}{r}  + \frac{2 \tilde{A}(r)}{r^3} + 2 \tilde{C}(r) + 2 \tilde{D}(r) r^2 \right) \\
v_{\theta} &= \cos \theta \left( - \frac{\tilde{B}(r)}{r} + \frac{\tilde{A}(r)}{r^3} - 2 \tilde{C}(r) - 4\tilde{D}(r) r \right)
\label{eq:flowDO}
\end{cases}
\end{equation}
Analytical expressions can be obtained for all coefficients  but  we report here only the expression for $\tilde{B}$: 
\begin{equation}
\tilde{B}(r) = B + \int_R^r \frac{1}{6} \tilde{f}(x) x^2 dx
\end{equation}
This is  the coefficient in front of the Stokeslet term, scaling as $1/r$, hence directly related to the force acting on the particle. 
As we discuss below, the diffusiophoretic velocity is
deduced from the force-free condition, which amounts to writing $\tilde{B}(r\rightarrow \infty) =0$.  
\subsection{Forces on the sphere}

\subsubsection{Pressure field and hydrodynamic force}
The pressure field $p$ can be computed from its full derivative
\begin{equation}
\mathrm{d} p = \partial_r p \, \mathrm{d}  r + \partial_{\theta} p \, \mathrm{d} \theta.
\end{equation} 
Using Eqs.~\eqref{eq:4flow2} and \eqref{eq:4flow3} we can integrate the pressure field and find
\begin{equation}
p = p_0 + \eta \cos \theta \partial_r \left[ \tilde{E}^2 F(r) \right]
\label{eq:pressureDO}
\end{equation}
The components of the hydrodynamic stress can be written as %on any centered sphere are 
\begin{equation}
\begin{cases}
\displaystyle \sigma_{rr} &= - p + 2 \eta \frac{\partial v_r}{\partial r} \\
\displaystyle \sigma_{r\theta} &= \eta \left( \frac{1}{r} \frac{\partial v_r}{\partial \theta} + \frac{\partial v_{\theta}}{\partial r} - \frac{v_{\theta}}{r} \right)
\end{cases}
\end{equation}
This leads to the expression of the normal and tangential hydrodynamic forces as
%so that 
%the normal force per unit area on the sphere, coming from hydrodynamic stresses, is therefore 
\begin{equation}
\frac{d f_r^{\rm hydro}}{dS} = \sigma_{rr}|_{r=R} = - p_0 - \eta \cos \theta \partial_{rrr} F(r) |_{r=R}
\label{eq:fosm1}
\end{equation}
and %The tangential force per unit area on the sphere, coming from hydrodynamic stresses is also simply
\begin{equation}
\frac{d f_{\theta}^{\rm hydro}}{dS} = \sigma_{r\theta}|_{r=R} = -\eta\frac{\sin \theta}{R} \partial_{rr} F(r) |_{r=R}
\label{eq:fosm2}
\end{equation}
where we took into account that derivatives in $F$ at  order 0 and 1 cancel at the sphere surface.

\subsubsection{Force from solute interaction}

In the force balance, we have also to take  into account the force exerted directly by the solute on the sphere via the interaction potential  $\mathcal{U}$. Because of the symmetry properties of $\mathcal{U}$, this force  has only a normal contribution. For a given unit spherical volume $d\tau = r^2 \sin \theta d\varphi d\theta dr$, this osmotic force writes
\begin{equation}
{d f_r^{\rm osm,(\tau)}} = - \eta \tilde{f}(r) \cos \theta \times d\tau.
\end{equation}
and the total osmotic force acting on a unit  surface $dS = R^2 \sin \theta d\theta d\varphi$ on the sphere is deduced as 
%the integral over the solute volume that stands in the $r$ direction over that unit surface
\begin{equation}
{d f_r^{\rm osm}}{} = - dS\times \eta \int_{R}^{R+\lambda}  \tilde{f}(r) \frac{r^2}{R^2} dr \cos \theta.
\label{eq:fosm3}
\end{equation}
 
\subsubsection{Total force on the sphere and diffusiophoretic velocity} 
 
The total force acting on the fluid is along the $z$ axis (the contribution on the perpendicular axis vanishes by symmetry) and
takes the expression
\begin{equation}
F_z = \int_{\theta = 0}^{\pi} \int_{\varphi = 0}^{2\pi} \left( d f^{\rm hydro}_r \cos \theta - d f^{\rm hydro}_{\theta} \sin \theta  + d f^{\rm osm}_r \cos \theta \right)
\end{equation}
This can be rewritten as
\begin{eqnarray}
&F_z&= -8\pi\eta \tilde{B}(R+\lambda) \nonumber \\
&&= - 6 \pi R \eta v_{DP} + 2 \pi R^2 \eta \int_R^{R+\lambda} \tilde{f}(r)\left( \frac{r}{R} - \frac{R}{3r} - \frac{2r^2}{3 R^2} \right) dr
\label{eq:ForceTot} 
\end{eqnarray}
Requiring that the total force on the sphere vanishes, $F_z = 0$, we then obtain 
\begin{equation}
v_{DP} = \frac{R}{3} \int_R^{R+\lambda} \tilde{f}(x) \left( \frac{x}{R} - \frac{R}{3x} - \frac{2x^2}{3 R^2} \right) dx 
\label{eq:4vDPf}
\end{equation}
Inserting the detailed expression of $\tilde{f}$, one gets
\begin{equation}
v_{DP} = \frac{R}{3\eta} \int_R^{R+\lambda} \frac{c(r,\theta) - c_0}{\cos\theta} \partial_r(-\mathcal{U})  \left( \frac{r}{R} - \frac{R}{3r} - \frac{2r^2}{3 R^2} \right) dr
\label{eq:4vDP}
\end{equation} 
%\begin{equation}
%
%\end{equation} 

\vspace{2mm}
\noindent \textit{Limiting expressions for a thin diffuse layer --}
%To recover the classical results for the diffusiophoretic velocity  of Anderson and co-workers, 
We now come back to the thin  diffuse layer regime where $\lambda \ll R$, which is the regime of interfacial flows. We need to prescribe the solute concentration 
profile $c(r,\theta)$ to calculate the diffusiophoretic velocity in Eq.~(\ref{eq:4vDP}). In the absence of external potential, the concentration verifies 
\begin{equation}
\begin{cases}
&\Delta c = 0 \\
&c(r \rightarrow \infty) =c_0 + \nabla c_{\infty} r \cos \theta \\
& \bm{\nabla} c (r = R) = 0 
\end{cases}
\end{equation}
and searching for a solution respecting the symmetry of the boundary conditions as $c(r,\theta) = c_0 + \nabla c_{\infty} R f(r) \cos \theta$, one finds 
\begin{equation}
c(r,\theta) = c_0 + R \nabla c_{\infty} \cos \theta  \left[ \frac{1}{2} \left( \frac{R}{r}\right)^2 + \frac{r}{R} \right]
\end{equation}
Now, in the presence of the external field ${\cal U}(r)$, 
one may simply extend the previous result by correcting the concentration profile  by the Boltzmann weight as
\begin{equation}
c(r,\theta) \simeq c_0 + R \nabla c_{\infty} \cos \theta  \left[ \frac{1}{2} \left( \frac{R}{r}\right)^2 + \frac{r}{R} \right] \exp\left( - \frac{\mathcal{U}}{k_B T}\right)
\end{equation}
which is a proper approximation in the regime of a small diffuse layer, valid both close to and far from the particle. 
In the limit where $\lambda \ll R$, one obtains  $c(R+x,\theta) = c_0 + \frac{3}{2} R \nabla c_{\infty} (1 + \frac{x^2}{R^2})  \cos \theta \exp\left( - \frac{\mathcal{U}}{k_B T}\right) + o(x^2)$ and this allows to simplify the diffusiophoretic velocity as
\begin{equation}
v_{DP} \simeq \frac{R^2 \nabla c_{\infty}}{2\eta} \int_0^{\lambda}  (1 + \frac{x^2}{R^2}  + o(x^2))   \exp\left( - \frac{\mathcal{U}}{k_B T}\right) \partial_x(-\mathcal{U})  \left( - \frac{x^2}{R^2}  + o(x^2) \right) dx
\end{equation}
%which gives easily enough
%\begin{equation}
%\begin{split}
%v_{DP} =  \frac{3}{2} c_{\infty} R \alpha \eta^{-1} & \bigg( \frac{1}{3} \int_0^{\lambda} e^{-\beta \mathcal{U}} \partial_x(-\mathcal{U}) ( 1 + \frac{x}{R} + \frac{x^2}{R^2}) dx \\
%& -  \frac{1}{9} \int_0^{\lambda} ( 1 - \frac{x}{R} + 2 \frac{x^2}{R^2})  e^{-\beta \mathcal{U}}  \partial_x(-\mathcal{U}) dx \\
%&-   \frac{2}{9} \int_0^{\lambda}  e^{-\beta \mathcal{U}} \partial_x(-\mathcal{U}) (1 + 2\frac{x}{R} + 2\frac{x^2}{R^2}) dx \bigg) \\
%v_{DO} = -  \frac{1}{2R} c_{\infty} \alpha \eta^{-1} &\int_0^{\lambda} e^{-\beta \mathcal{U}} \partial_x(-\mathcal{U}) x^2 dx \\
%v_{DO} =  - \frac{1}{2R} c_{\infty} \alpha \eta^{-1} &\bigg( \left[ \beta^{-1}(e^{-\beta \mathcal{U}}-1 ) x^2 \right]_{0}^{\lambda} \\
%& - 2 \int_0^{\lambda} \beta^{-1} \left( e^{-\beta \mathcal{U}} - 1 \right) x dx \bigg)
%\end{split}
%\end{equation} 
%And finally using that $\partial_z c_{\infty} = \nabla c_{\infty} = \frac{\alpha c_{\infty}}{R}$
yielding
\begin{equation}
v_{DP} = \nabla c_{\infty} \frac{k_B T}{\eta } \int_0^{\lambda} \left( e^{-\beta \mathcal{U}(x)} - 1 \right) x dx 
\end{equation}

With a similar reasoning one may also obtain
\begin{equation}
F_z = 6\pi \eta R (v_{DP}- v_{\rm slip}) = 6\pi \eta R v_{DP} - 6 \pi R k_B T \nabla c_{\infty} \int_0^{\lambda} \left( e^{-\beta \mathcal{U}(x)} - 1 \right) x dx 
\end{equation}
where $v_{slip}$ defines the osmotic contribution. Note that in the previous expressions, the upper limit $\lambda$ can now  safely be put to infinity: $\lambda\rightarrow \infty$.

\subsubsection{Local force on the diffusiophoretic particle}

From Eqs.~\eqref{eq:fosm1}-\eqref{eq:fosm3}, the total radial and tangential components of the local force on a surface element $dS = R^2 \sin \theta d\theta d\varphi$ are
\begin{equation}
df_r = - p_0 dS - dS \eta \left(+ \frac{3}{2R} v_{DP} - \int_R^{R+\lambda} \tilde{f}(x) \left( \frac{x}{2R} + \frac{R}{2x} - \frac{x^2}{R^2} \right) dx \right) \cos \theta
\end{equation}
and  
\begin{equation}
df_{\theta} = dS \eta \left(\frac{3}{2R} v_{DP} - \int_R^{R+\lambda} \tilde{f}(x) \left( \frac{x}{2R} - \frac{R}{2x} \right) dx    \right) \sin \theta.
\end{equation}
We can express $v_{DP}$ using Eq.~\eqref{eq:4vDPf} and this allows to write the local force in the compact form
\begin{equation}
\begin{cases}
\displaystyle d f_r = - p_0 dS + \frac{2}{3} \pi_s dS \cos \theta \\
\displaystyle d f_{\theta} = + \frac{1}{3} \pi_s dS \sin \theta
\end{cases}
\label{eq:localforce}
\end{equation}
where the local force is fully characterized by the {pressure} term
\begin{equation}
\pi_s = \eta \int_R^{R+\lambda} \tilde{f}(r) \left( \frac{R}{r} - \frac{r^2}{R^2} \right) dr
\end{equation}

It is interesting to express this pressure in the thin layer approximation:
\begin{equation}
\pi_s = - \frac{3}{2} R \nabla c_{\infty} \int_0^{\lambda}  (1 + \frac{x^2}{R^2}  + o(x^2))   \exp\left( - \frac{\mathcal{U}}{k_B T}\right) \partial_x(-\mathcal{U})  \left( 3 \frac{x}{R} + o(x^2) \right) dx 
\end{equation}
and we finally obtain the local force as
\begin{equation}
\begin{cases}
\displaystyle df_r &= - p_0 dS + 3 L_s k_B T \nabla c_{\infty} \cos \theta\, dS \\
\displaystyle df_{\theta} &= + \frac{3}{2} L_s k_B T \nabla c_{\infty} \sin \theta\, dS 
\end{cases}
\end{equation}
where 
\begin{equation}
L_s = \int_0^{\lambda} \left( e^{-\beta \mathcal{U}(x)} - 1\right) dx
\end{equation}
is a characteristic length scale of the interaction. We find in particular that $\Delta \Pi = L_s k_B T \nabla c_{\infty}$ is the relevant osmotic pressure, indicating that  
the relevant extension of the osmotic drop is the potential range $L_s$, and not the radius of the sphere $R$ as one may naively guess. 
%I find that it is ILLUMINATING to write the local force as computed by Anderson at this stage, which identifies to 
%\begin{equation}
%\colorboxed{red}{2 \eta v_{DO} R =  L_s k_B T \nabla c_{\infty} R^2 = \Delta \Pi R^2}
%\end{equation}
%Isn't that beautiful ? 

\section{Phoresis under electro-chemical gradients: general result and applications}

We generalize these calculations to the phoretic motion of a sphere under a gradient of electro-chemical potential. 

\subsection{Assumptions and variables}

The main working assumption here is that the perturbation to the electro-chemical potential $\mu$ is small, so that we may write
\begin{equation}
\begin{split}
\mu_i(r,\theta) &= k_B T \ln(\rho_i) +  V_i(r,\theta) \equiv \mu_{0,i}(r) + \tilde{\mu_i}(r,\theta)  \\
\rho_i &= \rho_{i,0}(r) + \tilde{\rho}_{i}(r,\theta) \\
V_i &= V_{0,i}(r) + \tilde{V}_{i}(r,\theta)
\end{split}
\end{equation}
where $i$ is the index of the solute specie, $\rho_i$ is the concentration of that specie, $V_i$ is the general potential acting on the specie (typically $V_i = q_i V_e + \mathcal{U}$ where $V_e$ is the electric potential, $q_i$ the charge of the specie, and $\mathcal{U}$ a neutral interaction potential). All quantities denoted as $y_0$ and $\tilde{y}$  correspond respectively to the equilibrium quantity and the perturbation under the applied field. In particular $\tilde{\mu}_i = k_B T \frac{\tilde{\rho}_i}{\rho_{0,i}} + \tilde{V}_i$. Equilibrium quantities only depend on the radial coordinate $r$ for symmetry reasons.

At equilibrium we have radial chemical equilibrium $\partial_r \mu_{0,i}=0$ and therefore
\begin{equation}
\rho_{0,i}(r) = c_0 \exp \left(-\frac{ V_{0,i}(r)}{k_B T}\right) 
\end{equation}
Additionally, Poisson's equation and the relevant electric boundary conditions allow to determine completely $\rho_0$ and $V_0$. 

In the presence of a small external field, we have the following linearized equation for the flux of specie $i$ 
\begin{equation}
\begin{split}
\bm{\nabla} \left( \frac{D_i}{k_B T} \tilde{\rho}_{i} (\bm{\nabla} V_0) + \frac{D_i}{k_B T} \rho_{0,i} ( \bm{\nabla} \tilde{V_i}) + D_i \bm{\nabla} \tilde{\rho}_{i} \right) = 0
\end{split}
\end{equation}
where $D_i$ is the diffusion coefficient of specie $i$. Since $ \nabla V_{i,0} = - k_B T\nabla  \rho_{0,i} / \rho_{0,i}$ we may simplify the first equation to
\begin{equation}
\bm{\nabla} \left(- D_i  \tilde{\rho}_{i}\bm{\nabla}  \rho_{0,i} / \rho_{0,i} + \frac{D_i}{k_B T}  \rho_{0,i}  (\bm{\nabla} \tilde{V_i}) + D_i \bm{\nabla} \tilde{\rho}_{i} \right) = 0 
\end{equation}
which simplifies to
%\begin{equation}
%\nabla \left(\rho_{0,i} \nabla  \left(   \frac{\tilde{\rho}_{i}}{\rho_{0,i}} + \frac{1}{k_B T} \tilde{V_i} \right) \right) = 0 
%\end{equation}
%or
\begin{equation}
\bm{\nabla} \left(\rho_{0,i} \bm{\nabla} \tilde{\mu_i} \right) = 0 
\label{eq:ChemPot}
\end{equation}

The applied field far from the particle surface is written in terms of a concentration or electric potential gradient, and $\tilde{\mu_i}\propto \nabla \mu_\infty$ the applied gradient of the electro-chemical potential. 
 %We suppose now that the perturbation in the field (either the concentration field or the force field) is a constant gradient far from the particle surface such that 
Due to the symmetry, one expects all perturbations to write as $\tilde{f}(r,\theta) = \tilde{f}(r) \cos\theta$ and
the $r$ dependence of the perturbation $\tilde{\mu_i}$ thus obeys the equation
\begin{equation}
\frac{1}{r^2}\frac{\partial}{\partial r} \left( r^2 \rho_{0,i} \frac{\partial \tilde{\mu}_{i}}{\partial r}  \right) - \frac{2}{r^2}  \tilde{\mu}_{i}(r) \rho_{0,i} = 0
\label{EOfluxGeneral}
\end{equation}

\subsection{Flow profile}
%Independently, we aim at simplifying the flow field. 
Going to the flow profile, the projection of the Stokes equation along  the unit vectors $\bm{e}_r$ and $\bm{e}_{\theta}$ leads to (following the same steps as in the previous section)
\begin{equation}
\begin{cases}
\displaystyle  \frac{\eta}{r^2 \sin \theta} \partial_{\theta} \left[ E^2 \psi \right] &= \partial_r p - \sum_i \rho_i \partial_r(-V_i)\\
\displaystyle \frac{- \eta}{r \sin \theta} \partial_r   \left[ E^2 \psi \right] &= \displaystyle \frac{1}{r}\partial_{\theta} p - \sum_i \rho_i \displaystyle \frac{\partial_{\theta}(-V_i)}{r}
\end{cases}
\end{equation}
From then on, and in order to simplify notations, we drop the sign corresponding to the sum over all particles.
We obtain to first order in the applied field %Now we use the notations above to write at linear order
\begin{equation}
\begin{split}
\frac{\eta}{r^2 \sin \theta} \partial_{\theta} \left[ E^2 \psi \right] &= \partial_r p - \rho_{0,i}(r) \partial_r(-V_{0,i})\\
&- \tilde{\rho}_i \partial_r(-V_{0,i}) \cos \theta - \rho_{0,i}(r) \partial_r(-\tilde{V_i}) \cos \theta    \\
\frac{- \eta}{r \sin \theta} \partial_r   \left[ E^2 \psi \right] &= \frac{1}{r}\partial_{\theta} p + \rho_{0,i} \frac{(-\tilde{V_i})}{r} \sin \theta
\end{split}
\end{equation}
Using the equilibrium distribution $- \partial_r V_{0,i} = k_B T \frac{\partial_r \rho_{0,i}}{\rho_{0,i}}$, one gets
\begin{equation}
\begin{split}
\frac{\eta}{r^2 \sin \theta} \partial_{\theta} \left[ E^2 \psi \right] &= \partial_r p - k_B T \partial_r \rho_{0,i}(r) \\
&- \tilde{\rho}_i  k_B T \frac{\partial_r \rho_{0,i}}{\rho_{0,i}} \cos \theta - \rho_{0,i}(r) \partial_r(-\tilde{V_i}) \cos \theta    \\
\frac{- \eta}{r \sin \theta} \partial_r   \left[ E^2 \psi \right] &= \frac{1}{r}\partial_{\theta} p + \rho_{0,i} \frac{(-\tilde{V_i})}{r} \sin \theta
\end{split}
\end{equation}
%Now we note that we may write $p' = p +  \tilde{V_i} \rho_{0,i} \cos \theta $ giving
%\begin{equation}
%\begin{split}
%\frac{\eta}{r^2 \sin \theta} \partial_{\theta} \left[ E^2 \psi \right] &= \partial_r p' - \rho_{0,i}(r) \partial_r(-V_0) \\
%&- \tilde{\rho}_i  k_B T \frac{\partial_r \rho_{0,i}}{\rho_{0,i}} \cos \theta - \partial_r \rho_{0,i}(r) \tilde{V_i} \cos \theta    \\
%\frac{- \eta}{r \sin \theta} \partial_r   \left[ E^2 \psi \right] &= \frac{1}{r}\partial_{\theta} p' 
%\end{split}
%\end{equation}
%Now we note that we can reabsorb in the pressure the component depending only on $r$ such that we write 
Introducing $p' = p - k_B T \rho_{0,i} +  \tilde{V}_i \rho_{0,i} \cos \theta $, one gets  the  compact formula
\begin{equation}
\begin{split}
\frac{\eta}{r^2 \sin \theta} \partial_{\theta} \left[ E^2 \psi \right] &= \partial_r p' - \tilde{\mu}_i(r) \partial_r \rho_{0,i}  \cos \theta \\
\frac{- \eta}{r \sin \theta} \partial_r   \left[ E^2 \psi \right] &= \frac{1}{r}\partial_{\theta} p' 
\end{split}
\label{EOSimpleGeneral}
\end{equation}
Eq.~\eqref{EOSimpleGeneral} has the exact same symmetries as Eq.~\eqref{eq:4flow2}, here with $\tilde{f}(r) = \frac{1 }{\eta}\tilde{\mu}_i(r) \partial_r \rho_{0,i}$. The flow profile therefore can be written as in Eq.~\eqref{eq:flowDO} and the pressure field is written similarly as in Eq.~\eqref{eq:pressureDO}  
\begin{equation}
p = p_0 + k_B T \rho_{0,i} - \tilde{V}_i \rho_{0,i} \cos \theta  + \eta \cos \theta \partial_r \left[ \tilde{E}^2 F(r) \right].
\end{equation}

\subsection{Phoretic velocity}

To simplify things, we consider first that there is no neutral potential. This contribution is easily added considering the previous section. 
To infer the phoretic velocity, we need to use the fact that the flow is force-less. For that, it is simple to write the total force acting on a large sphere of fluid say of radius $R_s \gg R + \lambda$ along the $z$ axis. The local hydrodynamic stresses write
\begin{equation}
\begin{split}
\frac{df_r^{hydro}}{dS}(R_s) = - p_0  & - k_B T \rho_{0,i} + \tilde{V}_i \rho_{0,i} \cos \theta \\
&+ \frac{3}{2}  \frac{\eta}{R_s^2} \left(- 3 R v_{P} + R^2 \int_R^{R+\lambda} \tilde{f}(r) \left( \frac{r}{R} - \frac{R}{3r} - \frac{2 r^2}{3R^2} \right) dr \right) \cos \theta 
\end{split}
\end{equation}
and  
\begin{equation}
\frac{df^{hydro}_{\theta}}{dS}(R_s) =  0
\end{equation}
and we note that $ \tilde{V}_i \rho_{0,i} (R_s) = + q_{+} \tilde{V} \rho_{0,+}(R_s)+ q_{-} \tilde{V}  \rho_{0,-}(R_s) = 0$ since the solution is uncharged far from the sphere. Also since the large sphere of radius $R_s$ is globally uncharged, the total force on the $z$ axis on this large sphere is therefore only the integral of the hydrodynamic stresses. Taking the condition that the flow is force-less we find a similar formula as in Eq.~\eqref{eq:4vDP} 
\begin{equation}
v_{P} = \frac{R}{3\eta} \int_R^{R+\lambda} \tilde{\mu}_i(r) \partial_r \rho_{0,i}  \left( \frac{r}{R} - \frac{R}{3r} - \frac{2r^2}{3 R^2} \right) dr
\label{eq:1vP}
\end{equation}
In Eq.~\eqref{eq:1vP} the potential $\tilde{\mu}_i(r)$ can be straightforwardly extended to account for both electric and neutral interactions.

\subsection{Local force balance}
The local force balance on the colloid is the sum of the hydrodynamic stresses and the electric force as
\begin{equation}
\begin{cases}
  \displaystyle \frac{df_r}{dS} &= -\displaystyle p_0  - k_B T \rho_{0,i} + \tilde{V}_i \rho_{0,i} \cos \theta +  \eta \frac{2}{3}   \int_R^{R+\lambda} f(r)  \left( \frac{R}{r} + \frac{r^2}{2R^2} \right) dr \cos \theta - \Sigma \partial_r V_e 
\\
  \displaystyle  \frac{df_{\theta}}{dS} \displaystyle &=   \displaystyle  \eta \frac{1}{3} \int_R^{R+\lambda} f(r)  \left(\frac{R}{r} - \frac{r^2}{R^2} \right) dr  \sin \theta - \frac{1}{R}\Sigma \partial_{\theta} V_e
\end{cases}
\label{eq:forcePG}
\end{equation}
where we used the expression for the phoretic velocity Eq.~\eqref{eq:1vP}. Eq.~\eqref{eq:forcePG} gives the expression of the local force balance in full generality. To simplify things further we assume a thin diffuse layer which allows to write 
\begin{equation}
\int_{R}^{R+\lambda} (q_i \rho_i) r^2 dr d^2\Omega (- \bm{\nabla} V_e(r)) = -R^2 \Sigma d^2 \Omega (- \bm{\nabla} V_e (R))
\end{equation}
where the main approximation here is  $(- \bm{\nabla} V_e(r)) \simeq  (- \bm{\nabla} V_e(R))$ and the rest is granted by electroneutrality. $d^2 \Omega$ is the solid angle on the sphere.
After a number of easy steps one finds
\begin{equation}
\begin{cases}
  \displaystyle \frac{df_r}{dS} &= -\displaystyle p_0  - k_B T \rho_{0,i}^{\infty} +  \eta \frac{2}{3}   \int_R^{R+\lambda} f(r)  \left( \frac{R}{r} - \frac{r^2}{R^2} \right) dr \cos \theta - \int_R^{R+\lambda} \frac{2r}{R^2} \rho_{0,i} \tilde{V}_i dr \cos\theta
\\
  \displaystyle  \frac{df_{\theta}}{dS} \displaystyle &=   \displaystyle  \eta \frac{1}{3} \int_R^{R+\lambda} f(r)  \left(\frac{R}{r} - \frac{r^2}{R^2} \right) dr  \sin \theta + \int_R^{R+\lambda} \frac{r}{R^2} \rho_{0,i} \tilde{V}_i dr \sin\theta
\end{cases}
\label{eq:forcePG1}
\end{equation}
Finally one remarks that terms in $ \int_R^{R+\lambda} \frac{r}{R^2} \rho_{0,i} \tilde{V}_i dr $ are of order $\lambda/R$ in front of the others, and therefore may be neglected in the thin layer approximation. Finally one arrives to the usual formulation, with the local force on a sphere surface element described by Eq.~\eqref{eq:localforce} and the pressure $\pi_s$ associated with the local force:
\begin{equation}
\pi_s  =  \int_R^{R+\lambda} \left( \frac{R}{r}  - \frac{r^2}{R^2} \right) \sum_i \tilde{\mu}_i(r) \partial_r \rho_{0,i} dr.
\end{equation}
%We are going to simplify $\pi_s$ now by integrating by parts and using the conservation of flux, but also the fact that the flux vanishes at the surface to get
%\begin{equation}
%\colorboxed{red}{\pi_s \frac{1}{R^2} = \rho_{0,i}(R+\lambda) \left(  \left( - \frac{R}{(R+\lambda)}  + \frac{(R+\lambda)^2}{R^2} \right) \tilde{\mu}_i(R+\lambda) - R \left( 1  + \frac{(R+\lambda)^3}{R^3} \right)  \partial_r\tilde{\mu}_i(R+\lambda)  \right)}
%\end{equation}
%\textcolor{red}{Before we move on we make here an important yet subtle statement on force reciprocity. In fact note that to derive Eq.~\eqref{eq:localforce}, we used the fact that at steady state and in the absence of external forces one can write that force exerted by the sphere on the fluid is the opposite of the force exerted by the fluid on the sphere. Here, in the potential presence of an external electric field, the statement is not so obvious. However, force reciprocity is still warrantied because overall the system is electroneutral and the electric field does not exert a force on the global system. }
%\textcolor{blue}{I'm still not super convinced by that. Because this equation assumes that local force balance holds and not global force balance.}

\subsection{Applications}

We now apply these results in various cases. 

\subsubsection{Application 1 : diffusiophoresis with neutral solute}

In the case of  diffusiophoresis with one neutral solute specie, one  has (using the notations above) $V_i = \mathcal{U}(r)$ and $\rho_i = c_{0} e^{- \mathcal{U}(r)/k_B T} + \tilde{\rho}$, with $\tilde{\rho}$ the perturbation under the external field. 
The local force thus writes
\begin{equation}
\pi_s =  \int_R^{R+\lambda} \left( \frac{R}{r}  - \frac{r^2}{R^2} \right) \frac{\tilde{\rho}}{\rho_{0,i}}  (- \partial_r \mathcal{U}) \rho_{0,i} dr
\end{equation}
Since $\tilde{\rho}(r) = \frac{c(r,\theta) - c_0}{\cos \theta}$,  one recovers the previous result in Eq.~\eqref{eq:fsDP}.

\subsubsection{Application 2 : electrophoresis in an electrolyte}

We consider the case of electrophoresis: namely a particle with a surface charge moving in an external applied electric field. Far from the particle the electric field is constant and reduces to the applied electric field, but it is modified (or screened) by the electrolyte solution close to the surface.
For simplicity we consider here two monovalent species, but the reasoning can be generalized easily. The local force on the particle is determined as 
\begin{equation}
\pi_s  =  \int_R^{R+\lambda} \left(  \frac{R}{r}  - \frac{r^2}{R^2} \right) \left(\tilde{\mu}_+(r) \partial_r \rho_{0,+} + \tilde{\mu}_-(r) \partial_r \rho_{0,-}\right) dr
\end{equation}
One can  simplify $\pi_s$ by integrating by parts $\rho_{0,\pm}$:
\begin{equation}
\pi_s = \left[\left( \frac{R}{r}  - \frac{r^2}{R^2} \right) \tilde{\mu}_{\pm} \rho_{0,\pm}  \right]_R^{R+\lambda} -  \int_R^{R+\lambda} \rho_{0,\pm}\partial_r \left[ \left(  \frac{R}{r}  - \frac{r^2}{R^2} \right) \tilde{\mu}_{\pm}(r) \right] dr
\end{equation}
%The term in brackets vanishes in $R$. Then we expand
%\begin{equation}
%\begin{split}
%\pi_s  =& \int_R^{R+\lambda} \rho_{0,\pm} \left( \frac{R}{r^2}  + \frac{2r}{R^2} \right) \tilde{\mu}_{\pm}(r) dr  -  \int_R^{R+\lambda} \rho_{0,\pm} \left( \frac{R}{r}  - \frac{r^2}{R^2} \right) \partial_r \tilde{\mu}_{\pm}(r) dr \\
%&+ \left[\left( \frac{R}{r}  - \frac{r^2}{R^2} \right) \tilde{\mu}_{\pm} \rho_{0,\pm}  \right]^{R+\lambda}
%\end{split}
%\end{equation}
%Now we rearrange a bit and integrate by parts only the second term
%\begin{equation}
%\begin{split}
%\pi_s \frac{1}{k_B T R} = &-  \int_R^{R+\lambda} \rho_{0,\pm} \left( \frac{R}{r^2}  + \frac{2r}{R^2} \right) \mathcal{U}_{\pm}(r) dr  -  \int_R^{R+\lambda}  \left( - \frac{R}{r^3}  + \frac{1}{R^2} \right)  \rho_{0,\pm} r^2 \partial_r \mathcal{U}_{\pm}(r) dr \\
%&+ \left[\left( - \frac{R}{r}  + \frac{r^2}{R^2} \right) \mathcal{U}_{\pm} \rho_{0,\pm}  \right]^{R+\lambda}
%\end{split}
%\end{equation}
%and integrate by parts only the second member
Rearranging the terms and integrating again by part, one obtains
\begin{equation}
\begin{split}
\pi_s  =& \int_R^{R+\lambda} \rho_{0,\pm} \left( \frac{R}{r^2}  + \frac{2r}{R^2} \right) \tilde{\mu}_{\pm}(r) dr  -  \int_R^{R+\lambda}  \left(  \frac{R}{2r^2}  + \frac{r}{R^2} \right) \partial_r \left( \rho_{0,\pm} r^2 \partial_r \tilde{\mu}_{\pm}(r) \right) dr \\
&+ \left[  \left(  \frac{R}{2r^2}  + \frac{r}{R^2} \right) \rho_{0,\pm} r^2 \partial_r \tilde{\mu}_{\pm}(r) \right]_R^{R+\lambda} + \left[\left( \frac{R}{r}  - \frac{r^2}{R^2} \right) \tilde{\mu}_{\pm} \rho_{0,\pm}  \right]^{R+\lambda}_R
\end{split}
\end{equation}
From Eq.~\eqref{EOfluxGeneral} we find that the integrals cancel each other and $\pi_s$ reduces to
\begin{equation}
\begin{split}
\pi_s = &- \frac{3R}{2} \rho_{0,\pm} (R) \partial_r \tilde{\mu}_{\pm}(R)  +   R \left(  \frac{1}{2}  + \frac{(R+\lambda)^3}{R^3} \right)  \rho_{0,\pm} (R+\lambda) \partial_r \tilde{\mu}_{\pm}(R+\lambda) \\
&+\left(  \frac{R}{R+\lambda}  - \frac{(R+\lambda)^2}{R^2} \right)\tilde{\mu}_{\pm}(R+\lambda) \rho_{0,\pm}(R+\lambda)
\end{split}
\end{equation}
Note that $\partial_r\tilde{\mu}_{\pm} (R)$ is actually the radial flux of particles at the boundary, and therefore is equal to $0$. Now we are interested in the far field expressions. {In this electrophoretic case}, one expects that there is no perturbation to the concentration field at distances beyond $R+\lambda$ ($\tilde{\rho} = 0$ and electroneutrality implies $\rho_{0,+} = \rho_{0,-}$). Therefore $\tilde{\mu}_{\pm}(R+\lambda) \sim \pm \frac{1}{k_B T} e \tilde{V}/\cos\theta$. In the far field, $\tilde{V}$ is simply $\tilde{V} = Er \cos\theta$ and we have $\tilde{\mu}_{\pm}(R+\lambda) \sim \pm \frac{1}{k_B T} e E \,(R+\lambda)$. As a consequence, the cation ($+$)  and anion ($-$) terms cancel in the above expression and one obtains the remarkable result:
\begin{equation}
\pi_s = 0
\end{equation}
In other words, no local surface force is applied on a particle undergoing electrophoresis. This is fully consistent with the expectations for the local force balance of
\cite{long1996simultaneous}.

\subsubsection{Application 3 : diffusiophoresis in electrolyte}

We now consider the case of diffusiophoresis in an electrolyte solution. For simplicity we take an electrolyte solution made of only one specie of monovalent anion and cation and identical diffusion coefficient. 
We  also perform the derivation in the Debye-H\"uckel limit, in order to obtain a tractable approximate result for the local force. 

\vspace{2mm}

\noindent \textit{Concentration profile -- } We consider first the equilibrium electrolyte profile in the absence of an external concentration field. The concentration profile obeys the simple Boltzmann equilibrium
\begin{equation}
\rho_{0,+}(r) = (c_0/2) \exp \left( - \frac{e V_0(r)}{k_B T}\right) \,\, \mathrm{and} \,\, \rho_{0,+}(r) = (c_0/2) \exp \left( + \frac{e V_0(r)}{k_B T}\right)
\end{equation}
and the potential $V_0(r)$ obeys the Poisson-Boltzmann equation
\begin{equation}
\Delta V_0 = \frac{c_0 e}{\epsilon} \sinh \left( \frac{e V_0}{k_B T}\right)
\label{eq:Poisson}
\end{equation}
where $\epsilon = \epsilon_0\epsilon_r$ is the permittivity of water. 
In the Debye-H\"uckel limit,  one linearizes the Poisson equation Eq.~(\ref{eq:Poisson})
%and using Gauss theorem we have 
%\begin{equation}
%\Delta V_0 \simeq \frac{c_0 e}{\epsilon} \frac{e V_0}{k_B T}
%\end{equation}
%the Debye length $\lambda_D^{-2} =  \frac{e^2 c_0}{\epsilon k_B T}$ we have
%\begin{equation}
%\Delta V_0 = \frac{V_0}{\lambda_D^2}
%\end{equation}
%This equation easily solves into
%\begin{equation}
%V_0 = \frac{C e^{-r/\lambda_D}}{r/\lambda_D}
%\end{equation}
%and using Gauss theorem we have the simple result
to obtain \begin{equation}
V_0(r) = \frac{\lambda_D \Sigma}{\epsilon} \frac{R}{R+\lambda_D} \frac{R}{r} e^{(R-r)/\lambda_D}
\end{equation}
where $\Sigma$ is the surface charge of the sphere and the Debye length is defined as $\lambda_D^{-2} =  \frac{e^2 c_0}{\epsilon k_B T}$. 

\vspace{2mm}

\noindent \textit{Chemical potential -- }
The chemical potential is obtained by solving perturbatively 
%We now look for the chemical potential. We are also going to perform an expansion of the chemical potential as a function of the electric potential. For that we solve iteratively 
Eq.~(\ref{eq:ChemPot}) as $\tilde{\mu} = \tilde{\mu}^{(0)} + \tilde{\mu}^{(1)} + ...$ where the expansion is in powers of the electrostatic potential
due to the particle, $eV_0/k_BT$.  The boundary condition at infinity writes
\begin{equation}
\tilde{\mu}_{\pm}(r \rightarrow \infty) = k_B T \frac{R\nabla c_{\infty} }{c_0} \frac{r}{R} \cos \theta
\end{equation}
To lowest order, one has $\bm{\nabla} \left(c_0 \bm{\nabla} \tilde{\mu_+}^{(0)} \right) = 0$ and therefore
\begin{equation}
\tilde{\mu}_+^{(0)} = k_B T \frac{R\nabla c_{\infty} }{c_0} \left( \frac{r}{R} + \frac{R^2}{2r^2} \right)
\end{equation}
using the no-flux boundary condition at the surface of the particle.
This is  similar to the result for diffusiophoresis with a neutral solute. 

For the next order one needs to solve
\begin{equation}
\bm{\nabla} \left(c_0 \bm{\nabla} \tilde{\mu}_+^{(1)} \right) = + \bm{\nabla} \left(c_0 \left( \frac{e V_0(r)}{k_B T} \right) \bm{\nabla} \tilde{\mu}_+^{(0)} \right)
\end{equation}
giving
%\begin{equation}
%\tilde{\mu_+}^{(1)} = \frac{r}{3} \int_{R}^r \frac{e V_0'(x)}{k_B T} \tilde{\mu_+}^{'(0)}(x) dx - \frac{1}{3r^2} \int_{R}^r x^3 \frac{e V_0'(x)}{k_B T} \tilde{\mu_+}^{'(0)}(x) dx + a r + b/r^2
%\end{equation}
%Where $a$ and $b$ Knowing that the flux should vanish at the surface sets $B = A R^3/2$. Then we have to pay attention to keep a finite flow at infinity. At infinity we have
%\begin{equation}
%\tilde{\mu_+}^{(1)} \rightarrow \frac{r}{3} \int_{R}^{\infty} \frac{e V_0'(x)}{k_B T} \tilde{\mu_+}^{'(0)}(x) dx  + A r 
%\end{equation}
%and therefore the value for $A =  - \frac{1}{3} \int_{R}^{\infty} \frac{e V_0'(x)}{k_B T} \tilde{\mu_+}^{'(0)}(x) dx$.So that we may rewrite the correction as
%\begin{equation}
%\tilde{\mu_+}^{(1)} = \frac{r}{3} \int_{\infty}^r \frac{e V_0'(x)}{k_B T} \tilde{\mu_+}^{'(0)}(x) dx - \frac{1}{3r^2} \int_{R}^r x^3 \frac{e V_0'(x)}{k_B T} \tilde{\mu_+}^{'(0)}(x) dx - \frac{1}{2}\frac{R^3}{3 r^2}\int_{R}^{\infty} \frac{e V_0'(x)}{k_B T} \tilde{\mu_+}^{'(0)}(x) dx
%\end{equation}
%Splitting the last integral in two parts
%\begin{equation}
%- \frac{R^3}{3 r^2}\int_{R}^{\infty} \frac{e V_0'(x)}{k_B T} \tilde{\mu_+}^{'(0)}(x) dx = - \frac{R^3}{3 r^2}\int_{R}^{r} \frac{e V_0'(x)}{k_B T} \tilde{\mu_+}^{'(0)}(x) dx - \frac{R^3}{3 r^2}\int_{r}^{\infty} \frac{e V_0'(x)}{k_B T} \tilde{\mu_+}^{'(0)}(x) dx
%\end{equation}
%yields the nicer formula
\begin{equation}
\begin{split}
\tilde{\mu}_+^{(1)}  = - \frac{R}{3} \left(\frac{r}{R} + \frac{R^2}{2 r^2}\right) \int_{r}^{\infty} \frac{e \partial_x V_0 (x)}{k_B T} \partial_x \tilde{\mu}_+^{(0)}(x) dx \\
 - \frac{R^3}{3r^2} \int_{R}^r \left( \frac{1}{2} + \frac{x^3}{R^3} \right) \frac{e \partial_x V_0(x)}{k_B T} \partial_x \tilde{\mu}_+^{(0)}(x) dx
\end{split} 
\end{equation}
where we used the no-flux boundary condition at the particle surface and also the condition of a bound  value for the chemical potential at infinity.

\vspace{2mm}

\noindent \textit{Local force on the surface --}
The expression for the local force acting on the sphere is written as 
\begin{equation}
\pi_s = +  \int_R^{R+\lambda} \left( \frac{R}{r}  - \frac{r^2}{R^2} \right) \left( \tilde{\mu}_+(r) \partial_r \rho_{0,+} + \tilde{\mu}_-(r) \partial_r \rho_{0,-}\right) dr 
\end{equation}
We expand the term in parenthesis as a function of $eV_0/k_B T$. At lowest order we get
\begin{equation}
\tilde{\mu}_+(r) \partial_r \rho_{0,+} + \tilde{\mu}_-(r) \partial_r \rho_{0,-} = \tilde{\mu}^{(0)}_+(r) c_0 \frac{-e\partial_r V_0}{k_B T} + \tilde{\mu}^{(0)}_-(r) c_0 \frac{e\partial_r V_0}{k_B T} + o\left(\frac{eV_0}{k_B T}\right)
\end{equation}
and  this order vanishes since $\tilde{\mu}^{(0)}_- = \tilde{\mu}^{(0)}_+$. Going to the next order we have
\begin{equation}
\begin{split}
\tilde{\mu}_+(r) \partial_r \rho_{0,+} + \tilde{\mu}_-(r) \partial_r \rho_{0,-} &= 2 \tilde{\mu}^{(0)}(r) c_0 \frac{e^2 V_0 \partial_r V_0}{(k_B T)^2} 
+ 2 \tilde{\mu}^{(1)}_+(r) c_0 \frac{-e\partial_r V_0}{k_B T} 
 + o\left(\left(\frac{eV_0}{k_B T}\right)^2\right)
\end{split}
\end{equation}
These terms may be formally integrated to calculate $\pi_s$. The expression for $\pi_s$ is cumbersome and  we do not report it here. Simpler forms are however obtained in some asymptotic regimes.  In the limit where the Debye length is small compared to the radius of the sphere $\lambda_D \ll R$ we get the approximated result
%Simply putting that into Mathematica yields a rather complicated result that we are however able to expand in the limit where the Debye length is small compared to the radius of the sphere: $\lambda_D \ll R$ and we find 
\begin{equation}
\pi_s(\lambda_D \ll R) \simeq \frac{9}{4} \frac{e^2 \Sigma^2 \nabla c_{\infty}  \lambda_D^3}{k_B T \epsilon^2}
\end{equation}
%Since 
%\begin{equation}
%\lambda_D^2 = \frac{\epsilon k_B T }{e^2 c_0} = \frac{\epsilon k_B T}{\sigma^2} \frac{\sigma^2(4\pi R^2)^2}{e^2} \frac{1}{(4\pi R^2)^2 c_0}
%\end{equation}
%we may rewrite
%\begin{equation}
%\pi_s = - \frac{9}{4} \frac{e^2 \nabla c_{\infty} R^2 c_0 \lambda_D^3}{\epsilon \lambda_D^2 \frac{e^2}{\sigma^2(4\pi R^2)^2} (4\pi R^2)^2 c_0^2}
%\end{equation}
%where we can recognize immediatly that $Q = \sigma(4\pi R^2)$ is the charge of the sphere and $Q_{D} = e \lambda_D  4\pi R^2 c_0$ is the compensating charge within the Debye layer. This ratio is not necessarily equal to 1 because this is not at all the meaning of the Debye layer, so in the end it's probably not the best way to write it. Finally using the expression for the Debye Layer one finds
%\begin{equation}
%\pi_s = - \frac{9}{4} k_B T \nabla c_{\infty} R^2 \lambda_D \frac{Q^2}{Q_D^2}
%\end{equation}
Introducing $\mathrm{Du} = \Sigma/e \lambda_D c_0$, a Dukhin number, the expression for $\pi_s$  can be rewritten as
\begin{equation}
\pi_s(\lambda_D \ll R) = \frac{9}{4} k_B T \nabla c_{\infty} \lambda_D \mathrm{Du}^2
\end{equation}
Gathering all contributions in concentration gives a scaling of $\pi_s \propto \nabla \left( 1/\sqrt{c}\right)$.
%Note that in the small debye length regime we expect $c$ to be rather large. 
%It also indicates what the correct length scale over which to measure the force is. 
This non-trivial dependence on the concentration differs from the  scaling of  the diffusiophoretic velocity, which scales as the gradient of the logarithm of the
concentration for diffusiophoresis with electrolytes.

% It doesn't make sense to include that here since we need a THIN LAYER HERE... This is however true because in that case there's no debate on the force balance
%Note that it also possible to find an approximation for the thick Debye layer
%\begin{equation}
%\pi_s(\lambda_D \gg R) \simeq \frac{1}{2} k_B T \nabla c_{\infty} \frac{R^2}{\lambda_D} \mathrm{Du}^2,
%\end{equation}
%in which case $\pi_s$ scales as $\pi_s \propto \nabla (\sqrt{c})$. 

\section{Diffusiophoresis of a porous sphere}

We consider now the case of diffusiophoresis of a porous sphere. This could also be considered as a minimal model for an entangled polymer. 
We will consider the case where the solute is neutral in order to simplify calculations. The calculations  could however be generalized to charged systems.
%For this we come back to the situation with no electric effects, but it is also possible to perform a similar derivation including general electric potentials and so on. 

\subsection{Flow profile}
Outside the sphere, for $r>R$, the flow profile is described by the Stokes equation, projected on the radial and tangential directions, see Eq.~\eqref{eq:StokesDP}. 
Inside the sphere, for $r<R$,  the Stokes equation now contains a supplementary Darcy term associated with the permeability of the sphere.
Projecting along $\bm{e}_r$ and $\bm{e}_{\theta}$ gives 
\begin{equation}
\begin{split}
\eta\left( \Delta v_r - \frac{2v_r}{r^2} - 2\frac{v_{\theta} \cos \theta}{r^2 \sin \theta} - \frac{2}{r^2} \frac{\partial v_{\theta}}{\partial \theta} \right)& - \frac{\eta}{\kappa} v_r = \partial_r p - c(r,\theta) \partial_r(-\mathcal{U}) \\
\eta \left( \Delta v_{\theta} - \frac{v_{\theta}}{r^2\sin^2\theta} + \frac{2}{r^2}\frac{\partial v_r}{\partial \theta}\right) & - \frac{\eta}{\kappa} v_\theta = \frac{1}{r}\partial_{\theta} p 
\end{split}
\end{equation}
where $\kappa$ is the permeability of the porous material, in units of a length squared. % $\mathrm{m^2}$.
For the porous sphere, the boundary conditions  at the sphere surface impose the continuity of the flow and stress. At infinity, the velocity should reduce 
to  $-v_{DP,p} \bm{e}_z$ in 
%are slightly different. The far field is moving in 
the reference frame of the particle. 
%as $-v_{DP,p} \bm{e}_z$ and the flow and stresses must be continuous at the porous interface. 
Using indices $1$ for inside the sphere and $2$ for outside, this gives
\begin{equation}
\begin{cases}
v_{2,r}(r \rightarrow \infty,\theta) =  - v_{DP,p} \cos \theta \,\,\, \mathrm{and} \,\,\,  v_{2,\theta}(r \rightarrow \infty,\theta) =   v_{DP,p} \sin \theta \\
v_{1,r}(r = R,\theta) = v_{2,r}(r=R,\theta) \,\,\, \mathrm{and} \,\,\, v_{1,\theta}(r = R,\theta) = v_{2,\theta}(r=R,\theta) \\
\sigma_{1,rr}(r=R,\theta) = \sigma_{2,rr}(r=R,\theta) \,\,\, \mathrm{and} \,\,\, \sigma_{1,r\theta}(r=R,\theta) = \sigma_{2,r\theta}(r=R,\theta)
\end{cases}
\end{equation}

We use a similar method as in Sec.~\ref{sec:DP}, defining a potential field $\psi = F(r) \sin^2\theta$ in each domain and operator $\tilde{E}$ such that 
\begin{equation}
\begin{cases}
\tilde{E}^4 F_1(r) - \frac{1}{\kappa} \tilde{E}^2 F_1(r) =  \tilde{f}(r) \\
\tilde{E}^4 F_2(r) =  \tilde{f}(r)
\end{cases}
\end{equation}
%\begin{equation}
%v_r = \frac{1}{r^2\sin\theta} \partial_{\theta} \psi \,\, \mathrm{and} \,\, v_{\theta} = - \frac{1}{r \sin \theta} \partial_r \psi
%\end{equation}
%We can rewrite Stokes as
%\begin{equation}
%\begin{split}
%\frac{\eta}{r^2 \sin \theta} \partial_{\theta} \left[ E^2 \psi \right] & - \frac{\eta}{\kappa} \frac{1}{r^2\sin\theta} \partial_{\theta} \psi  = \partial_r p - c(r,\theta) \partial_r(-\mathcal{U}) \\
%\frac{- \eta}{r \sin \theta} \partial_r   \left[ E^2 \psi \right] & + \frac{\eta}{\kappa} \frac{1}{r \sin \theta} \partial_r \psi= \frac{1}{r}\partial_{\theta} p 
%\end{split}
%\end{equation}
%This allows to simplify equations greatly and leads to 
%\begin{equation}
%\eta E^4 \psi = - \sin \theta \frac{\partial c(r,\theta)}{\partial \theta} \partial_r(-\mathcal{U})
%\end{equation}
%Considering the boundary conditions and the symmetry of $\mathcal{U}$ the only possibility is that $c(r,\theta) = c_{\infty} f(r) \cos \theta$, and therefore
%\begin{equation}
%\eta E^4 \psi - \frac{\eta}{\kappa} E^2 \psi=  \sin^2 \theta c_{\infty} f(r) \partial_r(-\mathcal{U})
%\end{equation}
%Now it is only fair to look for $\psi$ as $\psi = F(r) \sin^2\theta$ and we note that $E^2 \psi = \tilde{E}^2 F(r) \sin^2\theta$ where $\tilde{E}^2 = \frac{\partial^2}{\partial r^2} -  \frac{2}{r^2}$ such that
%\begin{equation}
%\tilde{E}^4 F(r) - \frac{1}{\kappa} \tilde{E}^2 F(r)=  \frac{c_{\infty}}{\eta} f(r) \partial_r(-\mathcal{U})
%\end{equation}
%For simplification we write $\tilde{f}(r) =  \frac{c_{\infty}}{\eta} f(r) \partial_r(-\mathcal{U})$ .
Outside the sphere, the general solution of this equation is
\begin{equation}
\begin{split}
F_2(r) = &\frac{A_2}{r} + B_2 r + r^2 C_2 + D_2 r^4- \frac{1}{r}\int_R^r \frac{\tilde{f}(x)x^4}{30}dx + r \int_R^r \frac{\tilde{f}(x)x^2}{6}dx \\
&- r^2 \int_R^r \frac{\tilde{f}(x)x}{6}dx + r^4 \int_R^r \frac{\tilde{f}(x)}{30x}dx
\end{split}
\end{equation} 
Inside the sphere, we introduce the following adjunct functions
\begin{equation}
\begin{split}
\alpha_a(r) &= \cosh(k_{\kappa} r)  - \frac{\sinh(k_{\kappa} r)}{k_{\kappa} r} \\
\alpha_b(r) &= \sinh(k_{\kappa} r)  - \frac{\cosh(k_{\kappa} r)}{k_{\kappa} r}
\label{alphas}
\end{split}
\end{equation}
where $k_{\kappa} = 1/\sqrt{\kappa}$ is the screening factor for the Darcy flow (inverse of a length). The solution inside the sphere thus writes
\begin{equation}
\begin{split}
F_1(r) =& \frac{A_1}{r} + r^2 C_1 + B_1\alpha_a(r) + D_1 \alpha_b(r) + \frac{1}{3r}\int_R^r \tilde{f}(x)x^2dx  \\
&- \frac{r^2}{3} \int_R^r \tilde{f}(x)x dx - \frac{\alpha_a(r)}{k_{\kappa}^3} \int_R^r \alpha_b(x) \tilde{f}(x) dx + \frac{\alpha_b(r)}{k_{\kappa}^3} \int_R^r \alpha_a(x) \tilde{f}(x) dx
\end{split}
\end{equation}

The integration constants $A_{1,2},... D_{1,2}$ are determined by the boundary conditions above. 
Also, the flow must be finite when $r\rightarrow \infty$, as well as when $r\rightarrow 0$. Note that the integrals do not diverge since $\tilde{f}$ is defined on a compact interval. Therefore we obtain  (for finite flow at infinity)
\begin{equation}
D_2 = - \int_R^{R+\lambda} \frac{\tilde{f}(x)}{30x}dx
\end{equation}
and we also have (for finite flow at small distances)
\begin{equation}
A_1 = \frac{1}{3}\int_0^R \tilde{f}(x) x^2 dx \,\,\, \mathrm{and} \,\,\, D_1 = \frac{1}{k_{\kappa}^3} \int_0^R \alpha_a(x) \tilde{f}(x) dx
\end{equation}
%but also (since $\alpha_b$ is also divergent) we get
%\begin{equation}
%D_1 = \frac{1}{k_{\kappa}^3} \int_0^R \alpha_a(x) \tilde{f}(x) dx
%\end{equation}
%Also we need the flow at infinity to be $-v_{DO} \bm{e}_z$ so this means $\psi(r\rightarrow \infty) = -(v_{DO}/2) r^2 \sin^2\theta$ therefore
The boundary condition at infinity yields
\begin{equation}
C_2 =  \int_R^{R+\lambda} \frac{\tilde{f}(x)x}{6}dx - {v_{DP,p}\over 2}
\end{equation}
The boundary conditions at the sphere surface impose continuity of $v_{\theta}(R)$, $v_r(R)$ and $\sigma_{rr}$ and $\sigma_{r\theta}$. The continuity of the velocities leads to the continuity of $F$ and $F'$ so that $F_1(R) = F_2(R)$ and $F'_1(R) = F'_2(R)$. The continuity of $\sigma_{r\theta}$ leads to the continuity of $F''$, and the continuity of $\sigma_{rr}$ to the continuity of the pressure. Some  straightforward calculations allow to show that the pressure takes the form
\begin{equation}
p_1 = p_0 + \eta \cos \theta \partial_r \left( \tilde{E}^2 F_1 - \frac{1}{\kappa} F_1 \right)
\end{equation}
such that the continuity of pressure amounts to
\begin{equation}
F_1'''(R)  - \frac{2}{R^2} F_1'(R) - \frac{\eta}{\kappa} F'_1(R) = F_2'''(R) - \frac{2}{R^2}F_2'(R)
\end{equation}
and because we already have $F'_1(R) = F'_2(R)$, we are left with
\begin{equation}
F_1'''(R) - \frac{1}{\kappa} F'_1(R) = F_2'''(R) 
\end{equation}
Altogether the boundary conditions are equivalent to the system of equations
\begin{equation}
\begin{cases}
F_1(R) &= F_2(R) \\
F_1'(R) &= F_2'(R) \\
F_1''(R) &= F_2''(R) \\
F_1'''(R) - \frac{1}{\kappa} F'_1(R) &= F_2'''(R) 
\end{cases}
\end{equation}
With 4 equations and 4 left undetermined integration constants, this system  allows us to completely calculate all left unknowns and determine the flow field. We do not report here  the full expressions for all constants, except for $B_2$ which is the prefactor of the Stokeslet  term  
%\begin{equation}
%\begin{split}
%B_2 = &-\frac{1}{6 \left( 3\alpha_a(R) (k_{\kappa}^2 R^2+1)- k_{\kappa}^2 R^2  \cosh( k_{\kappa} R) \right)} \bigg[ - 3k_{\kappa}^2 R^3 \alpha_a(R) \left(\int_{\lambda +R}^R x \tilde{f}(x) \, dx\right) \\
%   &+ 3 R^2 \left(\int_0^R \tilde{f}(x) \alpha_a(x) \, dx-3 k_{\kappa}^2 R v_{DP,p} \alpha_a(R) \right)- 3\alpha_a(R) \left(\int_0^R x^2 \tilde{f}(x) \, dx\right) \\
%   &+ R^3 \left( - 3 \alpha_a(R) + k_{\kappa}^2 R^2 \cosh (k_{\kappa} R)\right) \left(\int_{\lambda +R}^R \frac{\tilde{f}(x)}{x} \, dx\right) \bigg]
%\end{split}
%\end{equation}
%\begin{equation}
%\begin{split}
%B_2 = & \delta_{\kappa} \, \bigg( \frac{3R}{4}v_{DP,p} \,  -\frac{R^2}{12} \bigg[  \int_R^{R+\lambda} \left(\frac{3x}{R} + \frac{3}{k_{\kappa}^2 R^2} \frac{R}{x} - \frac{\cosh (k_{\kappa} R)}{\alpha_a(R)}\frac{R}{x} \right) \tilde{f}(x) \, dx \\
%   &+ 3 \frac{1}{k_{\kappa}^2R^2} \int_0^R \tilde{f}(x) \left(\frac{\alpha_a(x)}{\alpha_a(R)} -  \frac{x^2}{R^2} \right) \, dx  \bigg] \bigg)
%\end{split}
%\end{equation}
\begin{equation}
\begin{split}
B_2 = & \delta_{\kappa} \, \bigg( \frac{3R}{4}v_{DP,p} \,  -\frac{R^2}{12} \bigg[  \int_R^{R+\lambda} \left(\frac{3x}{R} - \frac{6}{k_{\kappa}^2 R^2} \frac{R}{x} - 3\frac{R}{x} + \frac{2 \cosh (k_{\kappa} R)}{\alpha_a(R)}\frac{R}{x} \right) \tilde{f}(x) \, dx \\
   &- 3 \frac{1}{k_{\kappa}^2R^2} \int_0^R \tilde{f}(x) \left(2 \frac{\alpha_a(x)}{\alpha_a(R)} +  \frac{x^2}{R^2} \right) \, dx \bigg] \bigg)
\end{split}
\end{equation}
where $\delta_{\kappa}$ is a dimensionless function characterizing the effect of porosity
\begin{equation}
\delta_{\kappa} = \left({ \frac{\cosh( k_{\kappa} R)}{\alpha_a( R)} + \frac{3}{2 (k_{\kappa} R)^2}}   \right)^{-1}
\label{eq:deltak}
\end{equation}
where the function $\alpha_a$ is defined in Eq.~(\ref{alphas}).
Note that $\delta_{\kappa} \rightarrow 1$ in the limit where the sphere is perfectly impermeable $\kappa \rightarrow 0$, allowing to recover the proper expression of $B$ as obtained for the plain sphere in Sec.~\ref{sec:DP}.
% I verified fully this equation. 

\subsection{Global force balance and diffusiophoretic velocity}

We define in a similar way as in Sec.~\ref{sec:DP}, $\tilde{B}(r) = B_2 + \frac{1}{6} \int_R^r \tilde{f}(r) r^2 dr$
and one may deduce the force from the asymptotic value for $\tilde{B}(r\rightarrow \infty)$:
%Now we want to know the total force on the porous sphere. \textcolor{red}{The total force on the sphere is the sum of the forces exerted by the fluid outside of the sphere (the fluid inside of the sphere does not contribute to force balance)} and therefore the calculation is exactly the same as in Sec.~\ref{sec:DP} and we have
\begin{equation}
F_z= - 8\pi\eta \tilde{B}(R+\lambda) = - 8\pi \eta B_2 - \frac{4 \pi}{3} \eta \int_R^{R+\lambda} \tilde{f}(r) r^2 dr
\end{equation}
Interestingly the viscous contribution to the force writes $F_{\rm hydro}=- 6\pi \delta_{\kappa}\, \eta R\, v_{DP,p}  $ with $\delta_{\kappa}$ defined in Eq.~\eqref{eq:deltak}.
%At this stage it is enlightening to write in a bit more detail the force balance using the expression for $\delta_{\kappa}$ in Eq.~\eqref{eq:deltak}
%\begin{equation}
%F_z =  - 6\pi  \eta R v_{DP,p}  \, \delta_{\kappa} + ... - \frac{4 \pi}{3} \eta \int_R^{R+\lambda} \tilde{f}(x) x^2 dx 
%\end{equation}
%where we have kept the characteristic Stoke's friction term and the osmotic contribution. Writing it this way it is clear that 
This indicates that $\delta_{\kappa}$ tunes the effective friction on the porous sphere. For any sphere permeability, we have $\delta_{\kappa} < 1$, and the effective friction is accordingly \textit{decreased} (therefore increasing {\it in fine} the diffusiophoretic velocity).
%the effect of the osmotic contribution. 
This effect is rather intuitive and is in agreement with the classical sedimentation of a porous sphere, where Stoke's friction is decreased as compared to the plain colloid case~\citep{sutherland1970sedimentation,joseph1964effect}. We will discuss further these results in the following subsections.

The motion is force-free $F_z = 0$ and one obtains the expression for $v_{DP,p}$:
%Since the sphere is moving with constant velocity $v_{DP,p}$, one enforce $F_z = 0$, resulting in an expression
%\begin{equation}
%\begin{split}
%v_{DP,p} = &\frac{1}{3 k_{\kappa}^2 R^3 \alpha_a(R)} \bigg( k_{\kappa}^2 R^2 \left( - \alpha_a(R) + \frac{1}{3} \cosh (k_{\kappa} R) \right) \int_R^{R+\lambda} x^2 \tilde{f}(x) \, dx \\
%& + k_{\kappa}^2 R^3 \alpha_a(R) \int_R^{R+\lambda} x \tilde{f}(x) \, dx \\
%&- \frac{R^3}{3} \left( k_{\kappa}^2 R^2 \cosh (k_{\kappa} R) - 3 \alpha_a(R) \right) \int^{R+\lambda}_R \frac{\tilde{f}(x)}{x} \, dx\\
%&+ R^2 \int_0^R \tilde{f}(x) \alpha_a(x)\, dx\\
%&- \alpha_a(R)  \int_0^{R+\lambda} x^2 \tilde{f}(x) \, dx \bigg)
%\end{split}
%\end{equation}
%
%I just want to write it a bit better so that it's more readable 
\begin{equation}
\begin{split}
v_{DP,p} = &\frac{R}{3} \int_R^{R+\lambda} \left(  \frac{r}{R} - \frac{R}{r}- \frac{1}{k_{\kappa}^2 R^2}\frac{r^2}{R^2}  -\frac{2}{k_{\kappa}^2 R^2} \frac{R}{r} \right) \tilde{f}(r) \, dr  \\ 
& + \frac{2 R}{9} \frac{\cosh (k_{\kappa} R)}{\alpha_a(R)} \int_R^{R+\lambda} \left(  \frac{R}{r} - \frac{r^2}{R^2} \right) \tilde{f}(r) \, dr \\
&+  \frac{R}{3 R^2 k_{\kappa}^2} \int_0^R \tilde{f}(r) \left( 2 \frac{\alpha_a(r)}{\alpha_a(R)} + \frac{r^2}{R^2} \right) \, dr 
\end{split}
\label{eq:vdpp}
\end{equation}
This equation can be rewritten in a compact form $v_{DP,p} = \frac{R}{3\eta} \int_0^{\infty} \frac{c(r,\theta) - c_0}{\cos\theta} \partial_r(-\mathcal{U})  \Phi(r) dr$ where the function $\Phi(r)$ takes the expression:
\begin{equation}
\begin{split}
\Phi(r) = & \mathbb{1}(r > R) \left(  \frac{r}{R} - \frac{R}{r}- \frac{1}{k_{\kappa}^2 R^2}\frac{r^2}{R^2}  -\frac{2}{k_{\kappa}^2 R^2} \frac{R}{r} + \frac{2}{3} \frac{\cosh (k_{\kappa} R)}{\alpha_a(R)} \left(  \frac{R}{r} - \frac{r^2}{R^2} \right) \right) \\
&+  \mathbb{1}(r < R) \frac{1}{ R^2 k_{\kappa}^2}  \left( 2 \frac{\alpha_a(r)}{\alpha_a(R)} + \frac{r^2}{R^2} \right)
\end{split}
\label{psiPhoresis}
\end{equation}

Taking the impermeable limit $\kappa \rightarrow 0$ (and thus  $k_{\kappa} \rightarrow \infty$) allows to recover the result of the non-porous sphere of Sec.~\ref{sec:DP}:
\begin{equation}
v_{DP,p}(\kappa = 0) =  R \int_R^{R+\lambda} \left( -\frac{2r^2}{9R^2}  + \frac{ r}{3 R} - \frac{R}{9r} \right)  \tilde{f}(r) \, dr \equiv v_{DP} 
\end{equation}
%
%And now, le test du diable, we take the limit when $\kappa \rightarrow 0$ and thus $k_{\kappa} \rightarrow \infty$. In this context we have $\alpha_a(x) \sim e^{k_{\kappa} x}/2$ and
%\begin{equation}
%\begin{split}
%v_{DO} =  &\frac{2}{3 k_{\kappa}^2 R^3 e^{k_{\kappa} R}} \bigg( k_{\kappa}^2 R^2 \left( -  e^{k_{\kappa} R}/2 + \frac{1}{3}  e^{k_{\kappa} R}/2 \right) \int_R^{R+\lambda} x^2 \tilde{f}(x) \, dx \\
%& - k_{\kappa}^2 R^3  e^{k_{\kappa} R}/2 \int_{R+\lambda}^R x \tilde{f}(x) \, dx \\
%&- \frac{R^3}{3} \left( k_{\kappa}^2 R^2  e^{k_{\kappa} R}/2 - 3  e^{k_{\kappa} R}/2 \right) \int^{\lambda +R}_R \frac{\tilde{f}(x)}{x} \, dx\\
%&+ R^2 \int_0^R \tilde{f}(x)  e^{k_{\kappa} x}/2 \, dx\\
%&- e^{k_{\kappa} R}/2  \int_0^{R+\lambda} x^2 \tilde{f}(x) \, dx \bigg)
%\end{split}
%\end{equation}
%Easily we see that $ e^{k_{\kappa} x-R}$ will go to 0 since $x < R$ and also we keep only highest order terms
%\begin{equation}
%\colorboxed{blue}{v_{DO} =   R \int_R^{R+\lambda} \left( -\frac{2x^2}{9R^2}  + \frac{ x}{3 R} - \frac{R}{9x} \right)  \tilde{f}(x) \, dx }
%\end{equation}
%Which is exactly what is expected from continuum standard dynamics. Now, the one million dollar question, what happens when $k_{\kappa}$ is not so infinite (what is the evolution of the DO velocity) ? Actually note that the other limit was almost computable... 
We can also expand for small permeabilities to get
\begin{equation}
v_{DP,p}(\kappa \rightarrow 0) = v_{DP} +  \frac{2 R}{9} \frac{1}{k_{\kappa} R} \int_R^{R+\lambda} \left( \frac{R}{r} - \frac{r^2}{R^2}  \right)  \tilde{f}(r) \, dr
\end{equation}
Working out the variations of the two terms one finds that the two geometrical contributions $\left( \frac{R}{r} - \frac{r^2}{R^2} \right)$ and $\left( -\frac{2r^2}{9R^2}  + \frac{ r}{3 R} - \frac{R}{9r} \right)$ are of the same sign (negative) for $r> R$. This means that the sphere porosity is \textit{increasing} the diffusiophoretic mobility. This effect is consistent with the reduction of friction and leads to a higher phoretic velocity. 
In the case of electrophoresis of porous particles and in the regime of a thin Debye-H\"uckel layer, a variety of behaviors are predicted and the effect of porosity is often entangled with other effects~\citep{hermans1955sedimentation,ohshima1994electrophoretic,huang2012electrophoretic}. The result is simpler for diffusiophoresis.

It is also interesting to explore the regime of a highly permeable sphere ($\kappa \rightarrow \infty$ or $k_{\kappa} \rightarrow 0$). In that case we find 
% all order terms in 1/(lambda_k R)^2 vanish and therefore only the terms in unity contribute. 
%Let's look at it. $\alpha_a(x) = \frac{k_{\kappa}^2 x^2}{3} + \frac{k_{\kappa}^4 x^4}{30}$ when $k_{\kappa} \rightarrow 0$ 
%\begin{equation}
%\begin{split}
%v_{DO,k_{\kappa} \rightarrow 0} = &\frac{1}{ k_{\kappa}^2 R^3 k_{\kappa}^2 R^2} \bigg( k_{\kappa}^2 R^2 \left( - \left( \frac{k_{\kappa}^2 R^2}{3} + \frac{k_{\kappa}^4 R^4}{30}\right) + \frac{1}{3} \left(1 + \frac{k_{\kappa}^2R^2}{2} + \frac{k_{\kappa}^4R^4}{24} \right) \right) \int_R^{R+\lambda} x^2 \tilde{f}(x) \, dx \\
%& + k_{\kappa}^2 R^3 \left( \frac{k_{\kappa}^2 R^2}{3} + \frac{k_{\kappa}^4 R^4}{30}\right) \int^{R+\lambda}_R x \tilde{f}(x) \, dx \\
%&- \frac{R^3}{3} \left( k_{\kappa}^2 R^2  \left(1 + \frac{k_{\kappa}^2R^2}{2} + \frac{k_{\kappa}^4R^4}{24} \right)  - 3 \left( \frac{k_{\kappa}^2 R^2}{3} + \frac{k_{\kappa}^4 R^4}{30}\right) \right) \int^{\lambda +R}_R \frac{\tilde{f}(x)}{x} \, dx\\
%&+ R^2 \int_0^R \tilde{f}(x)  \left( \frac{k_{\kappa}^2 x^2}{3} + \frac{k_{\kappa}^4 x^4}{30}\right)\, dx\\
%&- \left( \frac{k_{\kappa}^2 R^2}{3} + \frac{k_{\kappa}^4 R^4}{30}\right)  \int_0^{R+\lambda} x^2 \tilde{f}(x) \, dx \bigg)
%\end{split}
%\end{equation}
\begin{equation}
v_{DP,p}(\kappa \rightarrow \infty) = \frac{R}{(k_{\kappa}R)^2} \left( -  \int_R^{R+\lambda} \frac{r^2}{R^2} \tilde{f}(r) \, dr  +   \int_0^R \frac{r^2}{R^2}  \tilde{f}(r)  \, dr \right)
\end{equation}
The term in bracket can change sign  depending on the conditions and parameters and 
%This dependence is really interesting. In fact, as compared to the perfectly impermeable case, 
the velocity may accordingly reverse. %(especially if $\tilde{f}$ is very large very inside the sphere). 

%ICICICICI

%\vspace{2mm}
%The flow field outside the sphere is simply
%\begin{equation}
%v_{\theta} = \cos \theta \left( - \frac{\tilde{B}(r)}{r} + \frac{\tilde{A}(r)}{r^3} - 2 \tilde{C}(r) - 4\tilde{D}(r) r \right)
%\end{equation}
%where all coefficients may be explicited and
%\begin{equation}
%v_r = \sin \theta \left( 2 \frac{\tilde{B}(r)}{r}  + \frac{2 \tilde{A}(r)}{r^3} + 2 \tilde{C}(r) + 2 \tilde{D}(r) r^2 \right)
%\end{equation}

\subsection{Local surface force on the particle}
We now  compute the local force on the particle. 
%Since the boundary conditions are slightly different, the resulting forces are also slightly different. We have the following total radial component
%\begin{equation}
%p = p_0 + \eta \cos \theta \partial_r \left( \tilde{E}^2 F(r) \right)
%\end{equation}
%Then it is easy to compute the forces. The stresses are the following 
%\begin{align*}
%\sigma_{rr} &= - p + 2 \eta \frac{\partial v_r}{\partial r} \\
%\sigma_{r\theta} &= \eta \left( \frac{1}{r} \frac{\partial v_r}{\partial \theta} + \frac{\partial v_{\theta}}{\partial r} - \frac{v_{\theta}}{r} \right)
%\end{align*}
%
%
%For $\sigma_{rr}$ we have
%\begin{equation}
% \sigma_{rr}|_{r=R} = - p_0 - \eta \cos \theta \partial_{rrr} F(r) |_{r=R} - \frac{12}{R^3} \eta \cos \theta F(R)
%\end{equation}
%since order 1 derivatives in $F$ cancel at the sphere surface. 
%We also have the osmotic force contribution such that
The radial and tangential components take the following expressions in the present geometry:
\begin{equation}
\begin{split}
\frac{df_r}{dS} &= \sigma_{rr} - \int_R^{\infty} \eta \tilde{f}(r) \frac{r^2}{R^2} dr \cos\theta \\
&= - p_0 - \eta \cos \theta \partial_{rrr} F_2(r) |_{r=R} - 6 \eta \cos \theta \partial_r \left( \frac{F_2(r)}{r^2} \right)\bigg|_{r=R} - \int_R^{\infty} \eta \tilde{f}(r) \frac{r^2}{R^2} dr \cos \theta
\end{split}
\end{equation}
and  
%
%For the other component one finds
%\begin{equation}
% \sigma_{r\theta}|_{r=R} = -\eta\frac{\sin \theta}{R} \partial_{rr} F(r) |_{r=R} - 2 \eta \frac{\sin \theta}{R^3}F(R)
%\end{equation}
%such that 
\begin{equation}
\frac{df_{\theta}}{dS} = \sigma_{r\theta} = - 2 \eta \frac{\sin \theta}{R^3}F_2(R) + 2 \eta \frac{\sin \theta}{R^2} \partial_r F_2(r) |_{r=R}  -\eta\frac{\sin \theta}{R} \partial_{rr} F_2(r) |_{r=R} 
\end{equation}
The local forces hence write exactly as in Eq.~\eqref{eq:localforce1}, with the characteristic surface force $\pi_s$ as
%\begin{equation}
%\begin{split}
%\pi_s &= \frac{6}{k_{\kappa}^2 R^2}   \int_0^R \tilde{f}(r) \left(\frac{\alpha_a(r)}{ \alpha_a(R)} - \frac{r^2}{R^2} \right)\, dr\\
%& \left(3\left(\frac{2}{k_{\kappa}^2 R^2} +1\right) - 2 \frac{\cosh (k_{\kappa} R)}{ \alpha_a(R)} \right) \int_R^{R+\lambda} \left(\frac{R}{r} - \frac{r^2}{R^2} \right) \tilde{f}(r) \, dr  
%\label{eq:fsdpp}
%   \end{split}
%\end{equation}
%that we can rewrite fully as with
\begin{equation}
\begin{split}
\pi_s &= \frac{6}{k_{\kappa}^2 R^2}   \int_0^R \tilde{f}(r) \left(\frac{\alpha_a(r)}{ \alpha_a(R)} - \frac{r^2}{R^2} \right)\, dr\\
& + \left[3\left(\frac{2}{k_{\kappa}^2 R^2} +1\right) - 2 \frac{\cosh (k_{\kappa} R)}{ \alpha_a(R)} \right] \int_R^{R+\lambda} \left(\frac{R}{r} - \frac{r^2}{R^2} \right) \tilde{f}(r) \, dr  
\label{forcePsi}
   \end{split}
\end{equation}

%and
%\begin{equation}
%\begin{split}
%\frac{f_{\theta}}{dS} =&= -\frac{1}{3 k_{\kappa}^2}\bigg( 6  \int_0^R \tilde{f}(x) \left(\frac{x^2}{R^2} - \frac{\alpha_a(x)}{ \alpha_a(R)} \right)\, dx\\
%& ( - 2 k_{\kappa}^2 R^2 \frac{\cosh (k_{\kappa} R)}{ \alpha_a(R)} +3(2 +k_{\kappa}^2 R^2) ) \int_R^{\lambda +R} \left(\frac{x^2}{R^2} - \frac{R}{x}\right) \tilde{f}(x) \, dx   \bigg) \eta \cos \theta
%   \end{split}
%\end{equation}
%Note that it writes exactly as $f_r = 2 \pi_s \cos \theta$ and $f_{\theta} = \pi_s \sin \theta$. 
When the sphere is perfectly impermeable we easily recover the expression of Sec.~\ref{sec:DP}
\begin{equation}
\pi_s(\kappa = 0) = \int_R^{R+\lambda} \left( \frac{R}{r} -\frac{r^2}{R^2} \right) \tilde{f}(r) \, dr
\end{equation}
and going to the next order leads to 
\begin{equation}
\pi_s(\kappa \rightarrow 0) = \pi_s(\kappa = 0) \left( 1 - \frac{2}{k_{\kappa} R} \right)
\label{eq:psPorous}
\end{equation}
Porosity decreases friction and hence also the local force.

\subsection{Results in the thin diffuse layer limit}

In the thin diffuse layer limit, one may further approximate the previous results. 

{\it Concentration profile --} 
The concentration profile in the absence of the external potential verifies the Laplace equation together with boundary conditions
\begin{equation}
\begin{cases}
&\Delta c_1 = 0 \,\,\, \rm{for} \,\,\, r < R\\
&\Delta c_2 = 0 \,\,\, \rm{for} \,\,\, r > R\\
& c_2(r \rightarrow \infty) =c_0 + \nabla c_{\infty} r \cos \theta \\
& c_1(r = R) = c_1(r=R) \\
& D_1 \bm{\nabla} c (r = R) = D_2 \bm{\nabla} c (r = R) 
\end{cases}
\end{equation}
where the last equation represents conservation of flux at the porous interface; the indices $1$ and $2$ denote the solution inside and outside the sphere respectively. This set of equations is easily solved with the general form  $c_{1,2}(r,\theta) =  c_0 + \nabla c_{\infty} R f(r) \cos \theta$ (taking into account the fact the concentration profile should not diverge at the origin).

Now, in the presence of an external potential, one may 
%Adding in the Boltzmann weights in the presence of an interaction with the porous structure (as in Sec.~\ref{sec:DP}) we can 
approximate the concentration field by adding the Boltzmann weights (as in Sec.~\ref{sec:DP}):
\begin{equation}
\begin{cases}
c(r,\theta) \simeq \displaystyle c_0 + R \nabla c_{\infty}  e^{-\mathcal{U}(r)/k_B T} \left[ \frac{D_2 - D_1}{D_1 + 2D_2} \frac{R^2}{r^2} + \frac{r}{R} \right] \cos \theta \,\,\, &\mathrm{for} \,\,\, r > R \\
c(r,\theta) \simeq \displaystyle c_0 + R \nabla c_{\infty}  e^{-\mathcal{U}(r)/k_B T} \left[ \frac{3 D_2 }{D_1 + 2D_2}  \frac{r}{R} \right] \cos \theta \,\,\, &\mathrm{for} \,\,\, r < R
\end{cases}
\end{equation}
%And we are going to study this 
For a thin layer  $\lambda \ll R$, the concentration $c$ outside the sphere may be approximated as
\begin{equation}
c(R+x,\theta) = c_0 + \frac{3}{(D_1+2D_2)} R \nabla c_{\infty} e^{-\mathcal{U}(r)/k_B T} \left[ D_2 + D_1 \frac{x}{R} + (D_2 - D_1) \frac{x^2}{R^2}\right] \cos \theta
\end{equation}
%and now we insert that in the diffusiophoretic velocity and the local force expressions to get analytic insight. We will also look at the expansion when the sphere is almost impermeable $k_{\kappa} \rightarrow \infty$ to gain insight. 

%\subsubsection{diffusiophoretic velocity and local force}
{\it Diffusiophoretic velocity and local force --}
Performing expansions in $k_{\kappa} \rightarrow \infty$ allows to find
\begin{equation}
v_{DP,p}(\kappa \rightarrow 0)  = \frac{2D_2}{(D_1 + 2 D_2)} v_{DP}  \left( 1 + \frac{2 D_1}{D_2} \frac{1}{k_{\kappa} R}\right)  +  \frac{2}{ k_{\kappa} R}  \frac{D_2}{(D_1 + 2 D_2)} R L_s \nabla c_{\infty}  \frac{k_B T }{\eta}
\end{equation}
where we recall that
\begin{equation}
v_{DP} = \nabla c_{\infty} \frac{k_B T}{\eta } \int_0^{\lambda} \left( e^{- \mathcal{U}(x)/k_B T} - 1 \right) x dx 
\end{equation}
and  $L_s = \int_0^{\lambda} \left(e^{- \mathcal{U}(x)/k_B T} - 1\right) \, dx$.
%
%And now feeling lucky you can also try to write $\pi_s$ in this limiting regime 
%\begin{equation}
%\begin{split}
% \pi_s  &=  \frac{1}{k_{\kappa}^2 }\bigg( 6  \int_0^R \tilde{f}(x) \left(\frac{x^2}{R^2} - \frac{\alpha_a(x)}{ \alpha_a(R)} \right)\, dx\\
%& ( - 2 k_{\kappa}^2 R^2 \frac{\cosh (k_{\kappa} R)}{ \alpha_a(R)} +3(2 +k_{\kappa}^2 R^2) ) \int_R^{\lambda +R} \left(\frac{x^2}{R^2} - \frac{R}{x}\right) \tilde{f}(x) \, dx   \bigg) \eta
%\end{split}
%\end{equation}
%\begin{equation}
%\begin{split}
% \pi_s  &=  \frac{1}{k_{\kappa}^2 } ( - 2 k_{\kappa}^2 R^2 \frac{e^{k_{\kappa} R}/2}{ e^{k_{\kappa} R}/2}(1 + \frac{1}{k_{\kappa} R}) +3(2 +k_{\kappa}^2 R^2) ) \\ &\int_0^{\lambda} \left(\frac{R^2 + 2xR + x^2}{R^2} - \frac{R}{R}(1 - x/R + x^2/R^2)\right) \tilde{f}(x) \, dx   \eta
%\end{split}
%\end{equation}
%\begin{equation}
%\begin{split}
% \pi_s  &=  \left(1 - \frac{2}{k_{\kappa} R}\right) \int_0^{\lambda} \frac{3x}{R} \partial_x \mathcal{U} \frac{3 D_2}{D_1 + 2D_2} c_{\infty} \alpha e^{-\mathcal{U}(x)/k_B T} \, dx  
%\end{split}
%\end{equation}
%And now we integrate by parts leading to
%\begin{equation}
%\begin{split}
% \pi_s  &= - \left(1 - \frac{2}{k_{\kappa} R}\right) \frac{9 D_2}{D_1 + 2D_2} \nabla c_{\infty} \int_0^{\lambda} \left( e^{-\mathcal{U}(x)/k_B T} - 1\right) \, dx   
%\end{split}
%\end{equation}

The characteristic local force per unit surface can also be simply expressed as
\begin{equation}
\pi_s(\kappa \rightarrow 0) =  \pi_s(\kappa = 0) \frac{D_2}{D_2 + D_1/2} \left(1 - \frac{2}{k_{\kappa} R} \right)
\end{equation}
where we recall that $  \pi_s(\kappa = 0) = \frac{9}{2} L_s k_B T \nabla c_{\infty}$. We find that in any case the local surface force is \textit{decreased} as compared to the completely impermeable case.
Note that in the limit where the solute diffuses extremely slowly in the porous sphere, $D_1 \rightarrow 0$, it can be seen as impermeable to the solute and we recover Eq.~\eqref{eq:psPorous}.

\section{Summary and discussion}

%In the previous sections we established the presence of a significant local force on the surface of a diffusiophoretic particle. Let us emphasize that the situation is very different for electrophoretic transport. As was first discussed in \cite{long1996simultaneous}, for electrophoresis there is a {\it local} force balance between the direct electric force acting on the particle and the hydrodynamic shear acting on its surface:
%accordingly the local force simply vanishes identically ($\pi_s = 0$). In physical terms, this is due to the fact that the electric force acting on the colloid particle exactly balances the electrical force acting on the electric double layer because of local electroneutrality (the charge in the electric double layer being exactly opposite to the surface charge). This was exactly seen in our calculations. 

Our calculations allow to obtain an in-depth understanding of the local and global force balance obeyed by particles undergoing diffusiophoresis. 
While we considered in this paper the general situation of phoretic transport with neutral or charged solutes,  
we focus in this discussion on the results for diffusiophoresis. 

First, we showed that, at the global scale, the  force balance for  a particle moving under solute concentration gradients writes in a rather transparent form as
\begin{equation}
F= 6 \pi R \eta v_{DP} 
-  2 \pi R^2  \int_R^{\infty} c_0(r) (-\partial_r{\cal U})(r) \times \varphi(r) dr  \equiv 0
\label{FDObis}
\end{equation}
with $\varphi(r)=\frac{r}{R} - \frac{R}{3r} - \frac{2}{3} \left(\frac{r}{R}\right)^2 $ a dimensionless function, and the function $c_0(r)$ 
is proportional to the driving force, {\it i.e.} the solute concentration gradient far from the colloid: $c_0(r) \propto R \nabla c_\infty$. 
Eq.~(\ref{FDObis}) is the sum of the classic Stokes friction force on the sphere and a balancing force of osmotic origin, taking its root in the differential interaction $\mathcal{U}$ of the particle with the solute.
In the limiting case of a thin diffuse layer, the osmotic term simplifies to $ 6 \pi R k_B T \nabla c_{\infty} \int_0^{\infty} \left( e^{-\beta \mathcal{U}(z)} - 1 \right) z dz $ and the global force balance allows to recover the known expression for the diffusiophoretic velocity $v_{DP} = \nabla c_{\infty} \frac{k_B T}{\eta } \int_0^{\infty} \left( e^{-\beta \mathcal{U}(z)} - 1 \right) z dz$ \citep{anderson1991diffusiophoresis}. 
However the force balance in Eq.~(\ref{FDObis}) shows that one cannot simply predict the particle velocity by writing a balance between
the viscous term $6 \pi R \eta v_{DP} $ and a global osmotic force which would scale as $F_{osm} \sim R^2\times R\nabla \Pi$, with $\Pi = k_BT c_\infty$ the osmotic pressure. As discussed in the introduction, this estimate leads to a wrong prediction for the diffusiophoretic velocity, by a huge factor of order $(R/\lambda)^2$  where $\lambda$ is the size of the diffuse layer. This factor originates in the fact that the osmotic push takes its origin in the thin diffuse layer, and not at the scale $R$ of the particle. One has to account for the system dynamics at the scale of the diffuse layer in order to get a proper description of the osmotic transport. 
Discussions based on the naive force balance have led to considerable debates and misinterpretations of osmotically-driven transport of particles~\citep{cordova2008osmotic,julicher2009comment,fischer2009comment,PhysRevLett.103.079802,PhysRevLett.102.159802,brady2011particle,moran2017phoretic}. Our results fully resolve these concerns. 
%\begin{equation}
%v_{DP} = \frac{2 \pi R^2}{6\pi\eta R}   \int_R^{\infty} c_0(r) (-\partial_r{\cal U})(r) \times \varphi(r) dr 
%\label{VDPexactbis}
%\end{equation} 
%Remembering that $c_0(r) \propto R\nabla c_\infty$, this equation generalizes Eq.~(\ref{eq:PhoresisDOneutral})

Beyond the global force balance, a second outcome of our analysis concerns the {\it local} force balance. We have shown that 
particles undergoing phoretic transport experience a local force on their surface which takes the generic form
\begin{equation}
d {\bm f} = \left( - p_0 + \frac{2}{3} \pi_s \cos \theta \right) dS \, {\bm e_r}  + \left( \frac{1}{3} \pi_s \sin \theta \right)  dS  \, {\bm e_{\theta}} 
\label{eq:localforce1bis}
\end{equation}
where the local force is fully characterized by the force per unit area $\pi_s$ ($p_0$ is the bulk hydrostatic pressure and ${\bm e_r}$ and $ {\bm e_{\theta}}$ are unit vectors in the spherical coordinate system).

In the case of electrophoresis (with a thin diffuse layer), we have shown that $\pi_s$ vanishes identically: $\pi_s\equiv 0$. This simple and remarkable result is the consequence of the local electroneutrality which occurs for the \{particle + diffuse layer\}, so that the viscous and electric stresses balance each other locally. This result is in agreement with the seminal work of \cite{long1996simultaneous}.

In the case of diffusiophoresis however, the local force does not vanish. 
For a neutral solute and a thin diffuse layer, one gets the simple and transparent result 

\begin{equation}
\pi_s \simeq  \frac{9}{2}   k_B T L_s  \nabla c_{\infty}
\label{eqForceLocalebis}
\end{equation}
where 
$L_s = \int_R^{\infty} \left( e^{-\beta \mathcal{U}(x)} - 1\right) dx$ is a length  quantifying the excess adsorption of the solute on the sphere surface. 
This local force can be interpreted in simple terms. The osmotic force on the particle is actually  expected to scale as  $d\mathcal{V}_{\rm int}\times \nabla \Pi = d\mathcal{V}_{\rm int}  \nabla (k_B T c_{\infty})$ where $d\mathcal{V}_{\rm int} $ is the \textit{interaction volume}. In terms of the length $L_s$, which is the typical interaction lengthscale, one has $d\mathcal{V}_{\rm int} \approx  L_s\,dS$ and we recover the result of Eq.~(\ref{eqForceLocalebis}). Alternatively one
may realize that $\pi_s$ is of the order of the viscous surface stress and scales as $\pi_s \sim v_{DP} \eta/ \lambda$.
%leading accordingly to Eq.~\eqref{eqForceLocale} since $dS \simeq R^2$. 
We emphasize however this apparent simple reasoning is somewhat misleading and conceals the fact that a global force balance occurs at the scale of the particle leading to a zero force once integrated on the particle surface. 

\begin{figure}
\centering
 \includegraphics[width=0.95\textwidth]{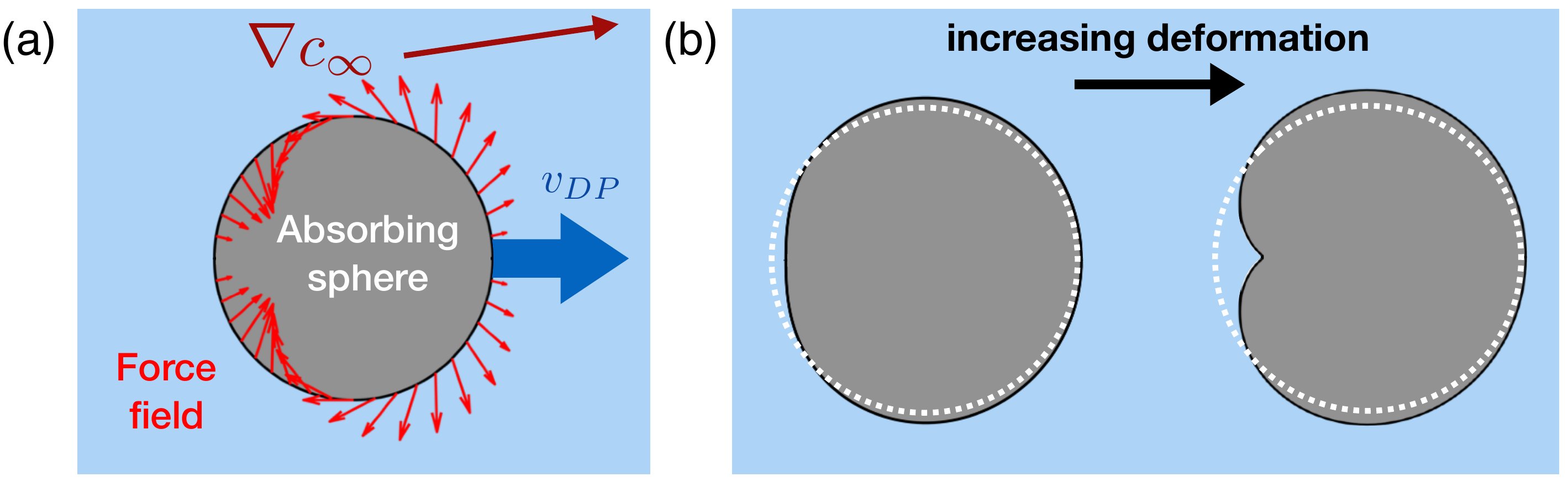}
 \caption{\textbf{Local force acting on a diffusiophoretic sphere.} (a) Local force field defined in Eq.~\eqref{eq:localforce1} acting on a sphere during diffusiophoresis with absorption at its surface in a solute gradient. The local force is plotted with an arbitrary amplitude factor (the same for each vector). (b) Resulting axisymmetric deformation of the sphere, when the deformation is assumed to be proportional to the local force, with an increasing amplitude from left to right. The dotted lines indicate the initial shape of the particle.}
 \label{fig:sphereForce}
\end{figure}

We have extended this result for the local force to a system of charged electrolytes, which in the limit of a thin Debye layer reduces to
\begin{equation}
\pi_s \simeq  \frac{9  \, Du^2}{4} k_B T\lambda_D \nabla c_{\infty}
\end{equation}
where $\lambda_D$  is the Debye length and $Du$ can be interpreted as a Dukhin number, here defined as $Du = \Sigma/e \lambda_D c_0$, where $\Sigma$ is the surface charge of the particle. This shows interestingly that the local osmotic push on the particle surface is a rather subtle combination of osmotic pressure and direct electric forces on the particle surface. 
% local force balance involves a rather subtle effectt of the surface charge and extension of the diffuse layer, which combine
%It is interesting for instance to comment the value obtained in the case of diffusiophoresis of a charged particle in a gradient of electrolyte. In that case the length scale of the interaction is characterized by the Debye length $\lambda_D$ ( writing $\lambda_D^{-2} =  \frac{e^2 c_0}{\epsilon k_B T}$). In the thin layer regime $\lambda_D \ll R$ one finds as expected that the relevant length of the osmotic push is $\lambda_D$ since  
%Notably the local force is modulated by the non-dimensional factor $Du = \Sigma/e \lambda_D c_0$, where $\Sigma$ is the surface charge of the particle. 
Another remark is that the local force, $\pi_s$, scales non-linearly with the electrolyte concentration, as 
$\pi_s \propto \nabla \left( 1/\sqrt{c}\right)$, and may induce a rather complex surface stress field on the particle surface. 
%which is rather surprising. Note that in the small debye length regime we expect $c$ to be rather large. This non-linearity reveals that deformation of the particles is "power-sensing" and therefore dependent on the absolute position of the particle. One may therefore expect similar spectacular behaviors as for diffusiophoretic "log-sensing"~\citep{palacci2012osmotic,shin2018cleaning}.
%The notable consequence of the presence of local forces on the sphere is that - provided it is soft - the sphere is expected to deform under these applied forces. In fact, 

Last but not least, the surface stresses in Eq.~(\ref{eq:localforce1bis}) generate an inhomogeneous local tension at the surface of the particle undergoing  diffusiophoresis. We plot in Fig.~\ref{fig:sphereForce}-a the corresponding force map. Accordingly, if one assumes that the particle may deform under a surface stress, this osmotic force field will induce a deformation of the particle. In Fig.~\ref{fig:sphereForce}-b, we sketch the deformation of a particle whose surface deforms elastically under a surface stress. 
We emphasize that this result is specific to diffusiophoresis and in strong contrast to the case of electrophoresis where the particle does {\it not} deform under the external field because of  the local electroneutrality as discussed above \citep{long1996simultaneous}. 

Let us estimate orders of magnitude for the deformation of a particle undergoing diffusiophoresis. We consider for simplicity a deformable droplet with radius $R$ and  surface tension $\gamma$: as a rule of thumb, the overall maximum deformation $\Delta R$ of the droplet is expected to scale as $\gamma R \Delta R \sim \pi_s R^2$.
%We now estimate the order of magnitude of the effect. 
Now one has typically the scaling $\pi_s \sim v_{DP} \eta/ \lambda$ in the thin diffuse layer limit. 
%Furthermore, the deformation of the particle $\Delta R/R$ is expected to be related to the local force by $\gamma R \Delta R \simeq \pi_s R^2$ where $\gamma$ is the surface tension of the particle. 
Therefore one expects $\Delta R/R \simeq v_{DP} \eta /(\gamma \lambda)$. 
%In \cite{lee2014osmotic}, the diffusio-osmotic flux in a microfluidic channel is measured to be $Q \simeq \mathrm{fL/min}$. Translating into a diffusio-osmotic velocity gives $v_{DO} \simeq 2 \mathrm{\mu m/s}$. As the diffusio-osmotic velocity scales like the diffusiophoretic velocity we have $v_{DP} \simeq v_{DO} \simeq 2 \mathrm{\mu m/s}$.  
Using typical values for the diffusiophoretic velocity $v_{DP}\sim 0.1 \mathrm{\mu}$m.s$^{-1}$ \citep{palacci2010colloidal}, surface tension $\gamma \approx 10. 10^{-3}$ N.m$^{-1}$~ \citep{pontani2012biomimetic}, fluid viscosity $\eta \approx 10^{-3}$ Pa.s and diffuse layer thickness, $\lambda \sim 10$ nm, then one predicts $\Delta R/R \sim 1$.
%we should take a smaller estimate, for example in a worst-case scenario $v_{DP} \simeq 20 \mathrm{nm/s}$.  
%Furthermore we consider the deformation of  \textit{e.g.} an oily droplet or a cell, and we take $\gamma \simeq 10 \mathrm{mN/m}$~ \citep{pontani2012biomimetic}. Choosing $\lambda \simeq 10 \mathrm{nm}$ and with $\eta \simeq 10^{-3} \mathrm{Pa.s}$ we find 
Large deformations are thus expected for the diffusiophoresis of droplets. We are not aware of an experimental study of this effect  for deformable particle undergoing diffusiophoresis. However, we note that in the context of thermo-phoresis, DNA molecules were reported to stretch under a temperature gradient ~\citep{jiang2007stretching}. Altough we did not explore thermophoretic transport in the present study, one may expect that similar surface stresses build up in this case. In a different context, a self-phoretic spherical cell with assymetric water pumps was predicted to substantially deform in a  rather similar way~\citep{yao2017numerical}. 

An interesting consequence of this deformation is that these effects may allow to separate deformable particles undergoing diffusiophoresis, for example if the deformation depends on the particle size. This would suggest to explore diffusiophoresis under solute gradients as an alternative (or complement) to separation techniques involving capillary electrophoresis, hence developing a capillary diffusiophoresis technique. 

%Our prediction of particle deformation under phoretic driving remains to be experimentally studied. We note however that in the context of thermo-phoresis, DNA was reported to stretch under a temperature gradient~\citep{jiang2007stretching}. Furthermore, in a different context, a self-phoretic spherical cell with assymetric water pumps was predicted to substantially deform in a similar way~\citep{yao2017numerical}. Furthermore, as we have mentioned earlier, with typical orders of magnitude one expects the deformation to be measurable with a typical deformation $\Delta R /R \simeq 20 \%$. Such effects could have interesting applications in the context of separation of particles, since their shape will differ depending on their size. 

\section*{Acknowledgements}
The authors would like to thank J\'er\'emie Palacci, Patrick Warren, Howard Stone, Daan Frenkel, Jasna Brujic and Lea-Laetitia Pontani for many interesting discussions.  L.B. acknowledges  support from European Union's Horizon 2020 Framework Program/European
Research Council Advanced Grant 785911 {\it Shadoks}.  This work was partially funded by Horizon 2020 program through 766972-FET-OPEN-NANOPHLOW.
S.M.  was supported partially by the MRSEC Program of the National Science Foundation under Award Number DMR-1420073.

\bibliographystyle{jfm}
%\bibliography{Diffusiophoresis}

\begin{thebibliography}{33}
\expandafter\ifx\csname natexlab\endcsname\relax\def\natexlab#1{#1}\fi
\def\au#1{#1} \def\ed#1{#1} \def\yr#1{#1}\def\at#1{#1}\def\jt#1{\textit{#1}}
  \def\bt#1{#1}\def\bvol#1{\textbf{#1}} \def\vol#1{#1} \def\pg#1{#1}
  \def\publ#1{#1}\def\arxiv#1{#1}\def\org#1{#1}\def\st#1{\textit{#1}}

\bibitem[Ab{\'e}cassis {\em et~al.\/}(2008)Ab{\'e}cassis, Cottin-Bizonne,
  Ybert, Ajdari \& Bocquet]{abecassis2008boosting}
{\sc \au{Ab{\'e}cassis, Benjamin}, \au{Cottin-Bizonne, C}, \au{Ybert, C},
  \au{Ajdari, A} \& \au{Bocquet, L}} \yr{2008}  \at{Boosting migration of large
  particles by solute contrasts}.  \jt{Nature materials}  \bvol{7}~(10),
  \pg{785}.

\bibitem[Ajdari \& Bocquet(2006)]{ajdari2006giant}
{\sc \au{Ajdari, Armand} \& \au{Bocquet, Lyd{\'e}ric}} \yr{2006}  \at{Giant
  amplification of interfacially driven transport by hydrodynamic slip:
  Diffusio-osmosis and beyond}.  \jt{Physical Review Letters}  \bvol{96}~(18),
  \pg{186102}.

\bibitem[Anderson(1989)]{anderson1989colloid}
{\sc \au{Anderson, John~L}} \yr{1989}  \at{Colloid transport by interfacial
  forces}.  \jt{Annual review of fluid mechanics}  \bvol{21}~(1),  \pg{61--99}.

\bibitem[Anderson \& Prieve(1991)]{anderson1991diffusiophoresis}
{\sc \au{Anderson, John~L} \& \au{Prieve, Dennis~C}} \yr{1991}
  \at{Diffusiophoresis caused by gradients of strongly adsorbing solutes}.
  \jt{Langmuir}  \bvol{7}~(2),  \pg{403--406}.

\bibitem[Brady(2011)]{brady2011particle}
{\sc \au{Brady, John~F}} \yr{2011}  \at{Particle motion driven by solute
  gradients with application to autonomous motion: continuum and colloidal
  perspectives}.  \jt{Journal of Fluid Mechanics}  \bvol{667},  \pg{216--259}.

\bibitem[C{\'o}rdova-Figueroa {\em et~al.\/}(2013)C{\'o}rdova-Figueroa, Brady
  \& Shklyaev]{cordova2013osmotic}
{\sc \au{C{\'o}rdova-Figueroa, UM}, \au{Brady, JF} \& \au{Shklyaev, S}}
  \yr{2013}  \at{Osmotic propulsion of colloidal particles via constant surface
  flux}.  \jt{Soft Matter}  \bvol{9}~(28),  \pg{6382--6390}.

\bibitem[C{\'o}rdova-Figueroa \& Brady(2008)]{cordova2008osmotic}
{\sc \au{C{\'o}rdova-Figueroa, Ubaldo~M} \& \au{Brady, John~F}} \yr{2008}
  \at{Osmotic propulsion: the osmotic motor}.  \jt{Physical Review Letters}
  \bvol{100}~(15),  \pg{158303}.

\bibitem[C\'ordova-Figueroa \& Brady(2009{\natexlab{{\em
  a\/}}})]{PhysRevLett.103.079802}
{\sc \au{C\'ordova-Figueroa, Ubaldo~M.} \& \au{Brady, John~F.}}
  \yr{2009{\natexlab{{\em a\/}}}}  \at{C\'ordova-figueroa and brady reply:}.
  \jt{Phys. Rev. Lett.}  \bvol{103},  \pg{079802}.

\bibitem[C\'ordova-Figueroa \& Brady(2009{\natexlab{{\em
  b\/}}})]{PhysRevLett.102.159802}
{\sc \au{C\'ordova-Figueroa, Ubaldo~M.} \& \au{Brady, John~F.}}
  \yr{2009{\natexlab{{\em b\/}}}}  \at{C\'ordova-figueroa and brady reply:}.
  \jt{Phys. Rev. Lett.}  \bvol{102},  \pg{159802}.

\bibitem[Derjaguin(1987)]{derjaguin1987some}
{\sc \au{Derjaguin, Boris~Vladimirovich}} \yr{1987}  \at{Some results from 50
  years' research on surface forces}.  \bt{In {\em Surface Forces and
  Surfactant Systems\/}},  \pg{pp. 17--30}.  \publ{Springer}.

\bibitem[Fischer \& Dhar(2009)]{fischer2009comment}
{\sc \au{Fischer, Thomas~M} \& \au{Dhar, Prajnaparamita}} \yr{2009}
  \at{Comment on “osmotic propulsion: the osmotic motor”}.  \jt{Physical
  Review Letters}  \bvol{102}~(15),  \pg{159801}.

\bibitem[Happel \& Brenner(2012)]{happel2012low}
{\sc \au{Happel, John} \& \au{Brenner, Howard}} \yr{2012} {\em Low Reynolds
  number hydrodynamics: with special applications to particulate media\/}, ,
  \vol{vol.~1}.  \publ{Springer Science \& Business Media}.

\bibitem[Hermans(1955)]{hermans1955sedimentation}
{\sc \au{Hermans, JJ}} \yr{1955}  \at{Sedimentation and electrophoresis of
  porous spheres}.  \jt{Journal of Polymer Science}  \bvol{18}~(90),
  \pg{527--534}.

\bibitem[Huang {\em et~al.\/}(2012)Huang, Hsu \& Lee]{huang2012electrophoretic}
{\sc \au{Huang, Cheng-Hsuan}, \au{Hsu, Hsuan-Pei} \& \au{Lee, Eric}} \yr{2012}
  \at{Electrophoretic motion of a charged porous sphere within micro-and
  nanochannels}.  \jt{Physical Chemistry Chemical Physics}  \bvol{14}~(2),
  \pg{657--667}.

\bibitem[Jiang \& Sano(2007)]{jiang2007stretching}
{\sc \au{Jiang, Hong-Ren} \& \au{Sano, Masaki}} \yr{2007}  \at{Stretching
  single molecular dna by temperature gradient}.  \jt{Applied Physics Letters}
  \bvol{91}~(15),  \pg{154104}.

\bibitem[Joseph \& Tao(1964)]{joseph1964effect}
{\sc \au{Joseph, DD} \& \au{Tao, LN}} \yr{1964}  \at{The effect of permeability
  on the slow motion of a porous sphere in a viscous liquid}.  \jt{ZAMM-Journal
  of Applied Mathematics and Mechanics/Zeitschrift f{\"u}r Angewandte
  Mathematik und Mechanik}  \bvol{44}~(8-9),  \pg{361--364}.

\bibitem[J{\"u}licher \& Prost(2009)]{julicher2009comment}
{\sc \au{J{\"u}licher, Frank} \& \au{Prost, Jacques}} \yr{2009}  \at{Comment on
  “osmotic propulsion: the osmotic motor”}.  \jt{Physical Review Letters}
  \bvol{103}~(7),  \pg{079801}.

\bibitem[Long {\em et~al.\/}(1996)Long, Viovy \& Ajdari]{long1996simultaneous}
{\sc \au{Long, Didier}, \au{Viovy, Jean-Louis} \& \au{Ajdari, Armand}}
  \yr{1996}  \at{Simultaneous action of electric fields and nonelectric forces
  on a polyelectrolyte: motion and deformation}.  \jt{Physical Review Letters}
  \bvol{76}~(20),  \pg{3858}.

\bibitem[Marbach \& Bocquet(2019)]{ReviewOsmosis}
{\sc \au{Marbach, Sophie} \& \au{Bocquet, Lyd{\'e}ric}} \yr{2019}  \at{Osmosis:
  from molecular insights to large-scale applications}.  \jt{Chemical Society
  Reviews} .

\bibitem[Michelin \& Lauga(2014)]{michelin2014phoretic}
{\sc \au{Michelin, S{\'e}bastien} \& \au{Lauga, Eric}} \yr{2014}  \at{Phoretic
  self-propulsion at finite p{\'e}clet numbers}.  \jt{Journal of Fluid
  Mechanics}  \bvol{747},  \pg{572--604}.

\bibitem[M{\"o}ller {\em et~al.\/}(2017)M{\"o}ller, Kriegel, Kie{\ss}, Sojo \&
  Braun]{moller2017steep}
{\sc \au{M{\"o}ller, Friederike~M}, \au{Kriegel, Franziska}, \au{Kie{\ss},
  Michael}, \au{Sojo, Victor} \& \au{Braun, Dieter}} \yr{2017}  \at{Steep ph
  gradients and directed colloid transport in a microfluidic alkaline
  hydrothermal pore}.  \jt{Angewandte Chemie International Edition}
  \bvol{56}~(9),  \pg{2340--2344}.

\bibitem[Moran \& Posner(2017)]{moran2017phoretic}
{\sc \au{Moran, Jeffrey~L} \& \au{Posner, Jonathan~D}} \yr{2017}  \at{Phoretic
  self-propulsion}.  \jt{Annual Review of Fluid Mechanics}  \bvol{49},
  \pg{511--540}.

\bibitem[Ohshima(1994)]{ohshima1994electrophoretic}
{\sc \au{Ohshima, Hiroyuki}} \yr{1994}  \at{Electrophoretic mobility of soft
  particles}.  \jt{Journal of colloid and interface science}  \bvol{163}~(2),
  \pg{474--483}.

\bibitem[Ohshima {\em et~al.\/}(1983)Ohshima, Healy \&
  White]{ohshima1983approximate}
{\sc \au{Ohshima, Hiroyuki}, \au{Healy, Thomas~W} \& \au{White, Lee~R}}
  \yr{1983}  \at{Approximate analytic expressions for the electrophoretic
  mobility of spherical colloidal particles and the conductivity of their
  dilute suspensions}.  \jt{Journal of the Chemical Society, Faraday
  Transactions 2: Molecular and Chemical Physics}  \bvol{79}~(11),
  \pg{1613--1628}.

\bibitem[Palacci {\em et~al.\/}(2010)Palacci, Ab{\'e}cassis, Cottin-Bizonne,
  Ybert \& Bocquet]{palacci2010colloidal}
{\sc \au{Palacci, J{\'e}r{\'e}mie}, \au{Ab{\'e}cassis, Benjamin},
  \au{Cottin-Bizonne, C{\'e}cile}, \au{Ybert, Christophe} \& \au{Bocquet,
  Lyd{\'e}ric}} \yr{2010}  \at{Colloidal motility and pattern formation under
  rectified diffusiophoresis}.  \jt{Physical review letters}  \bvol{104}~(13),
  \pg{138302}.

\bibitem[Palacci {\em et~al.\/}(2012)Palacci, Cottin-Bizonne, Ybert \&
  Bocquet]{palacci2012osmotic}
{\sc \au{Palacci, J{\'e}r{\'e}mie}, \au{Cottin-Bizonne, C{\'e}cile}, \au{Ybert,
  Christophe} \& \au{Bocquet, Lyd{\'e}ric}} \yr{2012}  \at{Osmotic traps for
  colloids and macromolecules based on logarithmic sensing in salt taxis}.
  \jt{Soft Matter}  \bvol{8}~(4),  \pg{980--994}.

\bibitem[Pontani {\em et~al.\/}(2012)Pontani, Jorjadze, Viasnoff \&
  Brujic]{pontani2012biomimetic}
{\sc \au{Pontani, Lea-Laetitia}, \au{Jorjadze, Ivane}, \au{Viasnoff, Virgile}
  \& \au{Brujic, Jasna}} \yr{2012}  \at{Biomimetic emulsions reveal the effect
  of mechanical forces on cell--cell adhesion}.  \jt{Proceedings of the
  National Academy of Sciences}  \bvol{109}~(25),  \pg{9839--9844}.

\bibitem[Sabass \& Seifert(2012)]{sabass2012dynamics}
{\sc \au{Sabass, Benedikt} \& \au{Seifert, Udo}} \yr{2012}  \at{Dynamics and
  efficiency of a self-propelled, diffusiophoretic swimmer}.  \jt{The Journal
  of chemical physics}  \bvol{136}~(6),  \pg{064508}.

\bibitem[Sharifi-Mood {\em et~al.\/}(2013)Sharifi-Mood, Koplik \&
  Maldarelli]{sharifi2013diffusiophoretic}
{\sc \au{Sharifi-Mood, Nima}, \au{Koplik, Joel} \& \au{Maldarelli, Charles}}
  \yr{2013}  \at{Diffusiophoretic self-propulsion of colloids driven by a
  surface reaction: the sub-micron particle regime for exponential and van der
  waals interactions}.  \jt{Physics of Fluids}  \bvol{25}~(1),  \pg{012001}.

\bibitem[Shin {\em et~al.\/}(2018)Shin, Warren \& Stone]{shin2018cleaning}
{\sc \au{Shin, Sangwoo}, \au{Warren, Patrick~B} \& \au{Stone, Howard~A}}
  \yr{2018}  \at{Cleaning by surfactant gradients: Particulate removal from
  porous materials and the significance of rinsing in laundry detergency}.
  \jt{Physical Review Applied}  \bvol{9}~(3),  \pg{034012}.

\bibitem[Sutherland \& Tan(1970)]{sutherland1970sedimentation}
{\sc \au{Sutherland, DN} \& \au{Tan, CT}} \yr{1970}  \at{Sedimentation of a
  porous sphere}.  \jt{Chemical Engineering Science}  \bvol{25}~(12),
  \pg{1948--1950}.

\bibitem[Velegol {\em et~al.\/}(2016)Velegol, Garg, Guha, Kar \&
  Kumar]{velegol2016origins}
{\sc \au{Velegol, Darrell}, \au{Garg, Astha}, \au{Guha, Rajarshi}, \au{Kar,
  Abhishek} \& \au{Kumar, Manish}} \yr{2016}  \at{Origins of concentration
  gradients for diffusiophoresis}.  \jt{Soft Matter}  \bvol{12}~(21),
  \pg{4686--4703}.

\bibitem[Yao \& Mori(2017)]{yao2017numerical}
{\sc \au{Yao, Lingxing} \& \au{Mori, Yoichiro}} \yr{2017}  \at{A numerical
  method for osmotic water flow and solute diffusion with deformable membrane
  boundaries in two spatial dimension}.  \jt{Journal of Computational Physics}
  \bvol{350},  \pg{728--746}.

\end{thebibliography}

%

\end{document}